\documentclass[modern]{aastex62}

\usepackage{amsmath}

\accepted{by ApJ}

\shorttitle{From the thermonuclear SN to the SNR}
\shortauthors{Ferrand et al.}

\begin{document}

\title{From the supernova to the supernova remnant:\\ the three-dimensional imprint of a thermonuclear explosion}

\correspondingauthor{Gilles Ferrand}
\email{gilles.ferrand@riken.jp}

\author[0000-0002-4231-8717]{Gilles Ferrand}
\newcommand{\ABBL}{Astrophysical Big Bang Laboratory (ABBL), 
RIKEN Cluster for Pioneering Research\\
Wak\={o}, Saitama, 351-0198 Japan}
\newcommand{\iTHEMS}{Interdisciplinary Theoretical and Mathematical Sciences Program (iTHEMS), RIKEN \\ 
Wak\={o}, Saitama, 351-0198 Japan}
\affiliation{\ABBL}
\affiliation{\iTHEMS}
\author[0000-0002-3222-9059]{Donald C. Warren}
\affiliation{\iTHEMS}
\author[0000-0002-0603-918X]{Masaomi Ono}
\affiliation{\ABBL}
\affiliation{\iTHEMS}
\author[0000-0002-7025-284X]{Shigehiro Nagataki}
\affiliation{\ABBL}
\affiliation{\iTHEMS}
\author[0000-0002-4460-0097]{Friedrich K.\ R{\"o}pke}
\affiliation{Zentrum f{\"u}r Astronomie der Universit{\"a}t Heidelberg, Institut f{\"u}r Theoretische Astrophysik, Philosophenweg 12, 69120 Heidelberg,
Germany}\affiliation{Heidelberger Institut f{\"u}r Theoretische Studien, Schloss-Wolfsbrunnenweg 35, 69118 Heidelberg, Germany}
\author[0000-0002-5044-2988]{Ivo R. Seitenzahl}
\affiliation{School of Science, University of New South Wales, Australian Defence Force Academy, Canberra, ACT 2600, Australia}


\begin{abstract}

Recent progress in the three-dimensional modeling of supernovae (SN) has shown the importance of asymmetries for the explosion. This calls for a reconsideration of the modeling of the subsequent phase, the supernova remnant (SNR), which has commonly relied on simplified ejecta models. In this paper we bridge SN and SNR studies by using the output of a SN simulation as the input of a SNR simulation carried on until 500~yr. We consider the case of a thermonuclear explosion of a carbon-oxygen white dwarf star as a model for a Type Ia SN; specifically we use the N100 delayed detonation model of Seitenzahl et al 2013. 
In order to analyze the morphology of the SNR, we locate the three discontinuities that delineate the shell of shocked matter: the forward shock, the contact discontinuity, and the reverse shock, and we decompose their radial variations as a function of angular scale and time. Assuming a uniform ambient medium, we find that the impact of the SN on the SNR may still be visible after hundreds of years. 
Previous 3D simulations aiming at reproducing Tycho's SNR, that started out from spherically symmetric initial conditions, failed to reproduce structures at the largest angular scales observed in X-rays. Our new simulations strongly suggest that the missing ingredient was the initial asymmetries from the SN itself. 
With this work we establish a way of assessing the viability of SN models based on the resulting morphology of the SNR.

\end{abstract}

\keywords{supernovae, supernova remnants}


\section{Introduction} 
\label{sec:intro}

Supernovae (SNe) are some of the most extreme events in the Universe. As endpoints of stellar evolution (of either massive stars or white dwarfs), they are important to understand the life cycle of stars in galaxies and play a crucial role in the synthesis of chemical elements. Despite substantial effort and recent progress, the very mechanisms by which stars explode have not been fully elucidated. For core-collapse supernovae (type II or Ib/c SNe), some self-consistent simulations have been able to unbind a star, although computational resources are insufficient to assess whether such models produce the required energetics in the end. It appears that multi-dimensionality, while challenging to simulate, is essential: instabilities play a key role in reviving the stalled shock inside the star (for recent reviews, see \cite{Janka2012, Janka2016}). For thermonuclear supernovae (type Ia SNe), it is still unknown whether they are produced by single-degenerate or double-generate progenitors -- or a mixture of both (see \cite{Hillebrandt2013} for a review). Yet, the (empirical) properties of their light curves have become a cornerstone of precision cosmology \citep{Riess1998,Perlmutter1999}.

The interaction between the stellar ejecta and the ambient medium generates a supernova remnant (SNR). Strong shock waves are produced, which convert the kinetic energy of the explosion into heat, supra-thermal particles, and magneto-hydrodynamic turbulence. The supersonic motion of the ejecta triggers a forward shock (FS) running into the interstellar medium (ISM), while the slowing down of the ejecta triggers a reverse shock (RS) running back into the ejecta. A shell of shocked matter forms, which is an X-ray emitter for typically thousands of years. The SNR is thus one of the ways to study the explosion mechanism, after the SN phase per se.
The multidimensional dynamics of SNRs have been studied using numerical simulations for some time \citep[e.g.][]{Chevalier1992, Blondin2001}. 
The SNR phase presents a hydrodynamic instability: the Rayleigh-Taylor instability (RTI) grows at the interface between the ejecta and the circumstellar medium, called the contact discontinuity (CD). 
\cite{Ferrand2010} and \cite{Warren2013} studied the evolution of a young SNR taking into account both the RTI at the CD and efficient particle acceleration at the FS. 
A~common trait of all these previous works is the use of idealized initial conditions. \cite{Ferrand2010} relied on the \cite{Chevalier1983} semi-analytical solutions for the shocked profiles in the early self-similar phase, while \cite{Warren2013} adopted an exponential ejecta profile following \cite{Dwarkadas1998}. Note that such initial conditions were effectively one-dimensional (1D), being functions of radius only, even though the subsequent SNR evolution was three-dimensional (3D). The main point of this new work is to use realistic initial conditions for the SNR phase. 

Commonly, researchers who study the evolution of SNRs do not simulate the supernova itself, and, conversely, researchers working on the explosion mechanism typically stop their simulations after several seconds or minutes. Our work fills this gap. 
Our postulate is that the initial stage of the SNR can reveal information about the SN explosion itself. Since 3D explosions are found to be asymmetric to some extent, we are looking for the imprint of the supernova on the observed morphology of the remnant, in terms of the spatial distribution of the ejecta. Recent studies that bridge SN physics and SNR physics, for the case of core-collapse explosions, are the works by Orlando et al on SN~1987A (\citeyear{Orlando2015}) and on Cas~A (\citeyear{Orlando2016}). In both simulations, a 1D SN model was mapped to 3D before computing the SNR evolution. In the latter one, asymmetries were added in the ejecta, so as to match the asymmetries of the observed SNR. This demonstrates the relevance of morphological studies to investigate the SN to SNR connection. 
Our approach is to start with a 3D SN model as obtained from state-of-the art simulations, and compute the hydro evolution from there. In this paper we consider only the thermonuclear case. It is known that, statistically, Type Ia SNRs tend to be more spherical and symmetric than core-collapse SNRs \citep{Lopez2011}. However their ejecta also exhibit complex structures, and we show in this work that this leads to observable features in the remnant structure. For the case of core-collapse SNe, we refer the reader to the preliminary reports by \cite{Ellinger2013} and \cite{Gabler2016}.

The structure of our paper is as follows. In Section~\ref{sec:method} we summarize the SN simulation used and how we proceed for the SNR simulation, we also explain how we analyze the 3D structure of the evolving SNR. In Section~\ref{sec:results} we show the development over the first 500 years of a Tycho-like SNR, and the imprint of the initial SN conditions. In Section~\ref{sec:discussion} we discuss the reliability of our results, and comment on implications regarding observations. In Section~\ref{sec:conclusion} we summarize and present the next steps.

\section{Method}
\label{sec:method}

In this section we describe the numerical simulation, from the SN to the SNR, including tools for the analysis of the morphology.

\subsection{SN model}
\label{sec:SN}

We treat the case of a thermonuclear explosion. For this first work connecting SN and SNR simulations, we concentrate on the single degenerate scenario, that is one accreting white dwarf, and we consider the most popular explosion model, a delayed detonation \citep{Khokhlov1991}. One of the questions regarding thermonuclear explosions is whether the explosion propagates as a sub-sonic front (deflagration), as a supersonic front (detonation), or as a deflagration that at some point(s) transitions to a detonation (the deflagration-to-detonation model, known as DDT).
A~set of 2D DDT models recovers the range of luminosities, spectra, and colors of SNe~Ia \citep{Blondin2011}, 3D DDT models show reasonable agreement with the spectra of normal SNe~Ia although they fall short to recover the width-luminosity relation and colors \citep{Sim2013}.
The newer 3D simulations were performed with updated methods \citep{Ciaraldi-Schoolmann2013}. While persisting shortcomings can potentially be attributed to the modeling approaches, it is also possible that they point to a general failure of Chandrasekhar-mass models to explain normal SNe~Ia (sub-Chandrasekhar mass explosions seem to be a promising alternative, e.g. \cite{Sim2010,Shen2018}). In this work we explore whether the simulated ejecta structure of a 3D DDT explosion model translates into observable features in the remnant phase, which allows to further constrain this scenario. We note that a DDT model is also favored by \cite{Williams2017} from observations of Tycho's SNR. 

\cite{Seitenzahl2013} presented a suite of 14 three-dimensional, high-resolution hydrodynamical simulations of a DDT explosion of a Chandrasekhar-mass white dwarf. The ignition is an important yet hard to simulate process, so it was parametrized via different choices of ignition kernel locations. 
We pick the N100 model, which is the most promising one for reproducing the typical brightness of normal SNe~Ia by producing about half a solar mass of $^{56}$Ni. 
In the \cite{Seitenzahl2013} simulations nucleosynthesis was computed at runtime only for the few species that significantly contribute to the energetics. A full nuclear reaction network, including 196 isotopes for 30 elements from Z = 1 to 32, was run in post-processing, using the technique of tracer particles. The results of the post-processing were then remapped from a particle representation back to a grid of resolution $200^3$. We use these data cubes. 
The SN simulation was run until a time of about $100\, \mathrm{s}$. By that time the evolution is observed to be essentially self-similar \citep{Ropke2005}.\footnote{We note that things would be different for the core-collapse case, where it could take a week for the ejecta to settle.} The radial extent of the ejecta is $\simeq 2.9\times 10^{11}$ cm, that is $9.4 \times 10^{-8}$ pc or $5$ solar radii. The total kinetic energy of the ejecta is $E_\mathrm{SN} = 1.4\times 10^{51}$ erg, for a total mass $M_\mathrm{ej} = 1.4 M_\odot$, implying a maximal speed of about 28,000 km\,s$^{-1}$. 

\begin{figure}[ht!]
\centering
\includegraphics[width=1\textwidth]{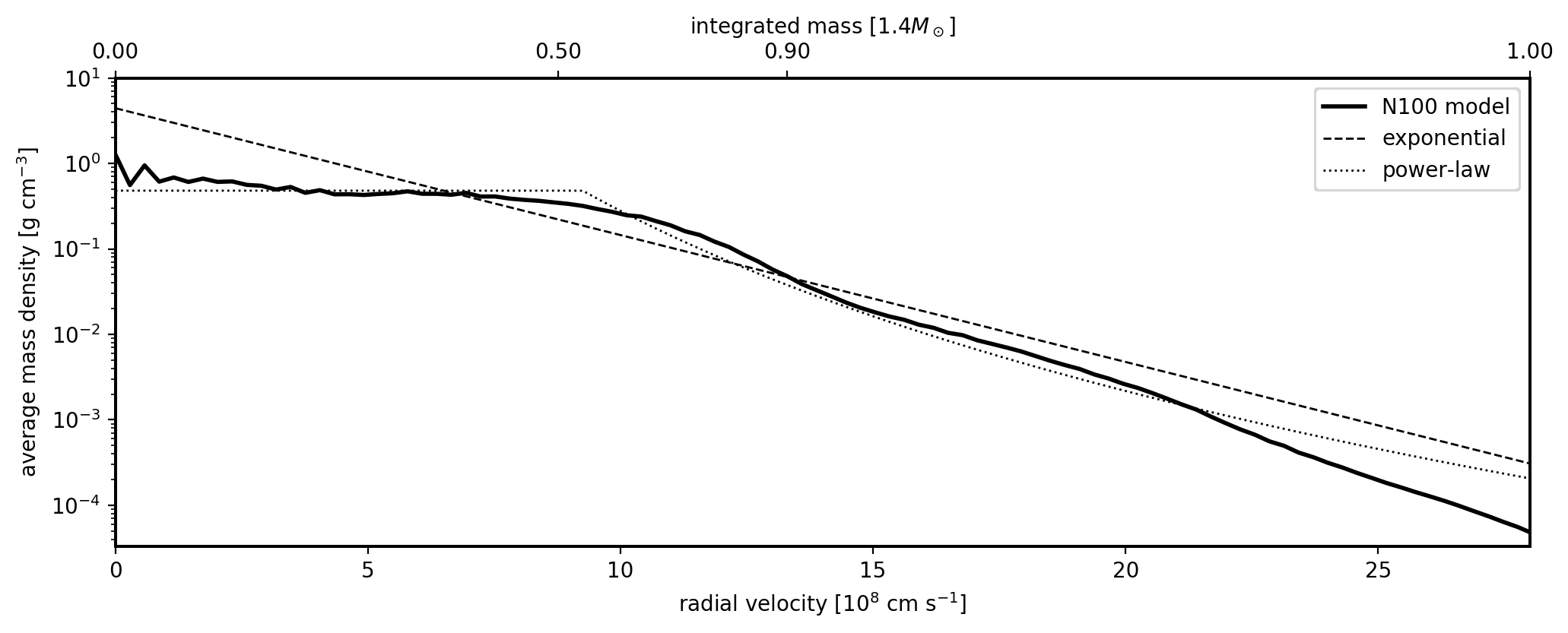}
\caption{\label{fig:rho-1D}
Average radial profile of the mass density of the N100 model at $t=100$~s (thick solid line). Thinner lines are analytical profiles with the same mass and energy: an exponential (dashed, \cite{Dwarkadas1998}), and a truncated power-law of index $n=7$ (dotted, \cite{Decourchelle1994}).}
\end{figure}

\begin{figure}[ht!]
\centering
\includegraphics[width=1\textwidth]{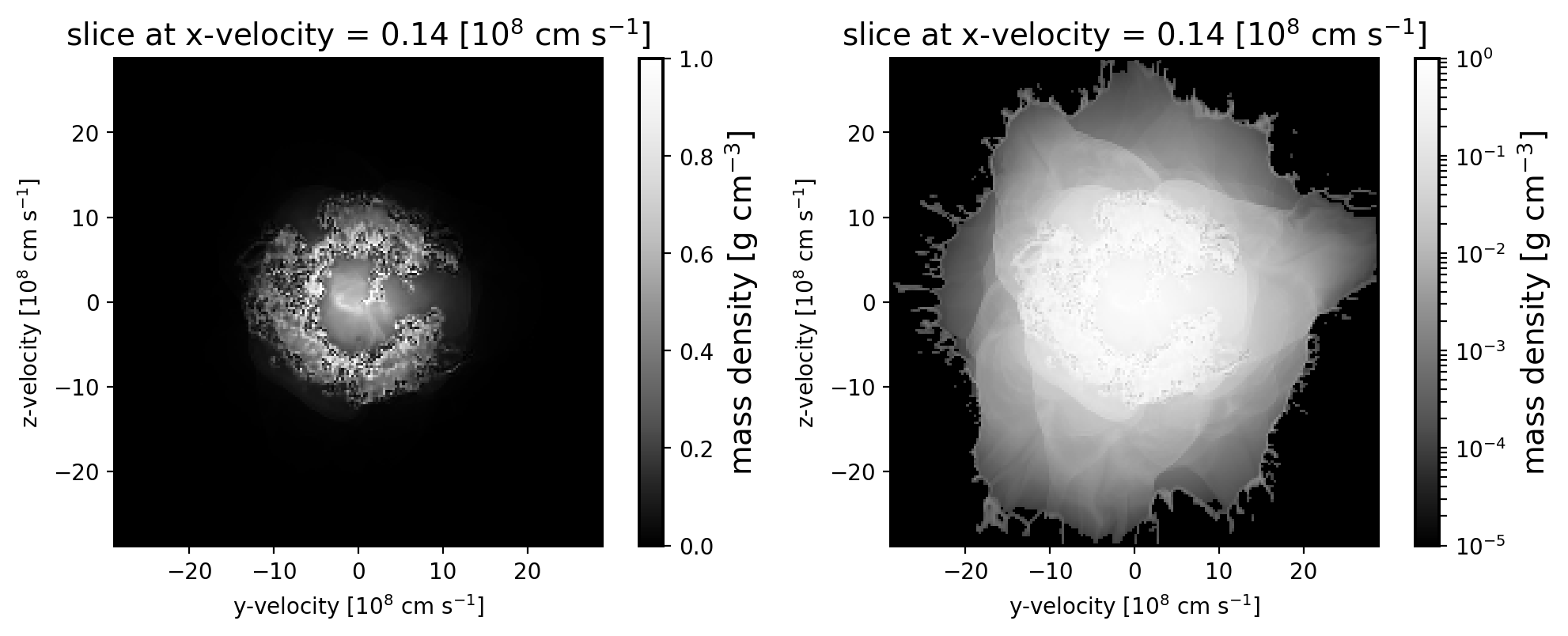}
\caption{\label{fig:rho-2D}
Slice of the mass density of the N100 model at $t=100$~s, taken at the centre of explosion along the $x$-axis. Left: the linear scale emphasizes the central part of the ejecta; Right: the logarithmic scale reveals the outer envelope of the ejecta. Animations showing the slicing of the data cube along the 200 bins of the $x$-axis are available online. 
}
\end{figure}

The 3D structure produced by the SN simulation is substantially different from a 1D averaged profile. The 1D radial profile of the mass density, after averaging over all angles, is shown in Figure~\ref{fig:rho-1D}, compared with the exponential and power-law profiles typically used in the SNR community. One can see that it may be approximated by a 2-step exponential profile. 2D slices of the mass density field are shown in Figure~\ref{fig:rho-2D}.\footnote{Interactive 3D models are hosted online on Sketchfab at \url{https://skfb.ly/6pKYW}. These show iso-contours, of the mass density, and of the abundance of three of the main tracer species (${}^{12}\mathrm{C}$, ${}^{16}\mathrm{O}$, and ${}^{56}\mathrm{Ni}$). These iso-contours, as well as volume renderings of the full data cubes, are  part of an immersive science exhibit developed at the Astrophysical Big Bang Laboratory at RIKEN, that can be experienced using room-scale virtual reality (VR) hardware \citep{Ferrand2018}.} They reveal the complex morphology of the ejecta. Most of the mass is concentrated in the central part, with a compact nickel core, and irregular outer shells of carbon and oxygen. The small-scale structures in the centre are the result of the deflagration phase, while the sharper outer interfaces were produced by multiple detonation fronts.

\vspace{1mm}
\subsection{SNR evolution}
\label{sec:SNR}

To simulate the SNR phase, we perform hydro simulations in a way similar to \cite{Ferrand2010} and following papers (\citeyear{Ferrand2012,Ferrand2014}). We review here the main features of the simulations, concentrating on those aspects that have been updated or are treated differently. 
We are using a custom version of the code RAMSES \citep{Teyssier2002}, an Eulerian code on a Cartesian grid. Even though spherical coordinates may feel like a natural choice to describe a SN or SNR, we are precisely interested here in deviations from spherical symmetry. Our computational domain is however spherical, with boundary conditions applied in the corners of the cube. 
We further comment on the grid geometry and resolution in Section~\ref{sec:disc-num}.
We use a grid of size $256^3$, which contains the entire SN data cube obtained after remapping. New compared to our previous SNR papers, we simulate the entire remnant, rather than an octant, since we no longer assume any symmetry.

\paragraph{Initial and boundary conditions.}

\begin{figure}[t]
\centering
\includegraphics[width=0.85\textwidth]{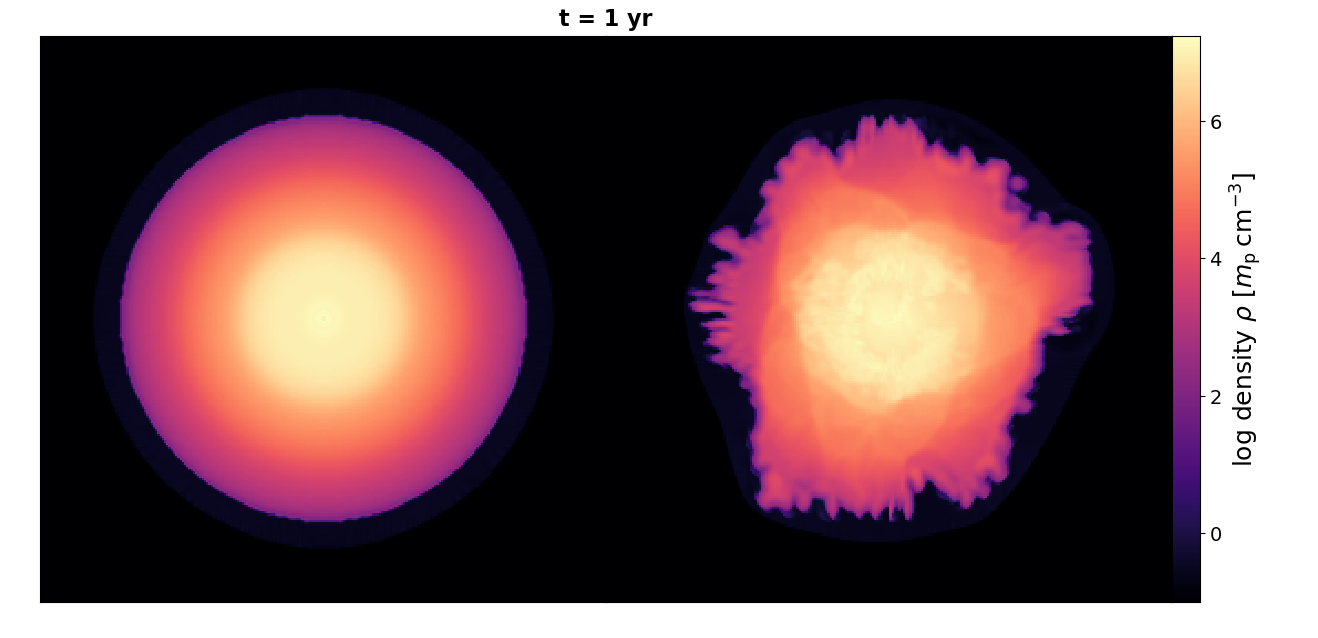}
\includegraphics[width=0.85\textwidth]{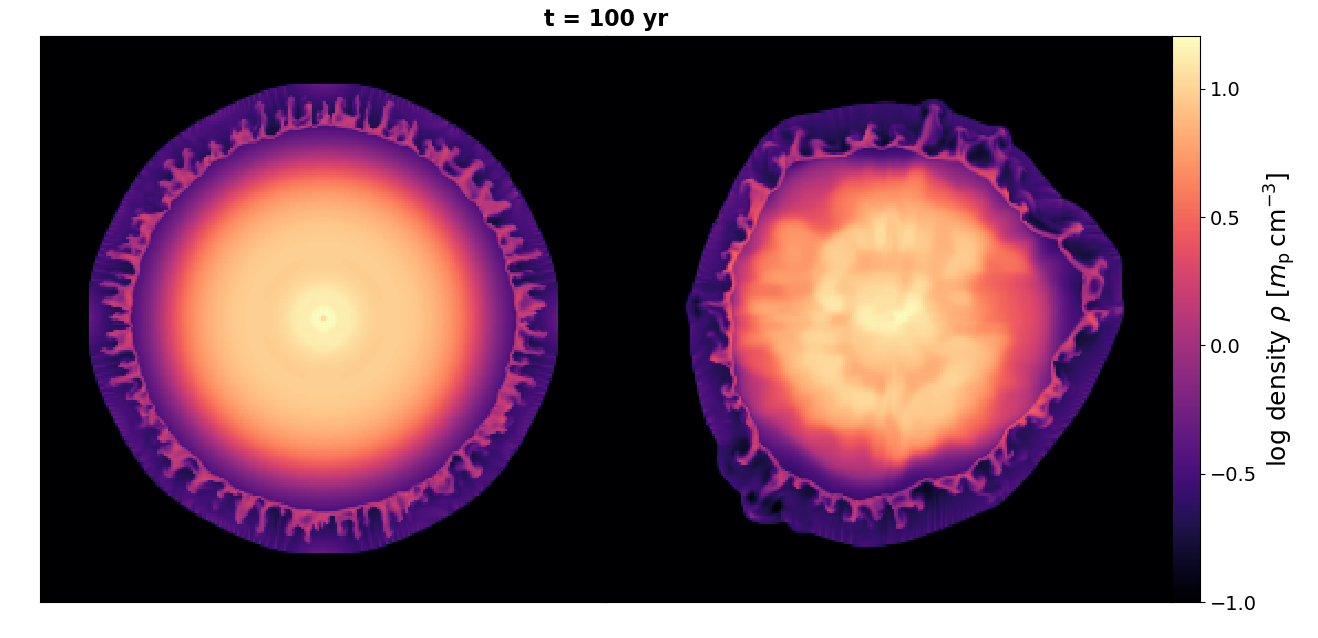}
\includegraphics[width=0.85\textwidth]{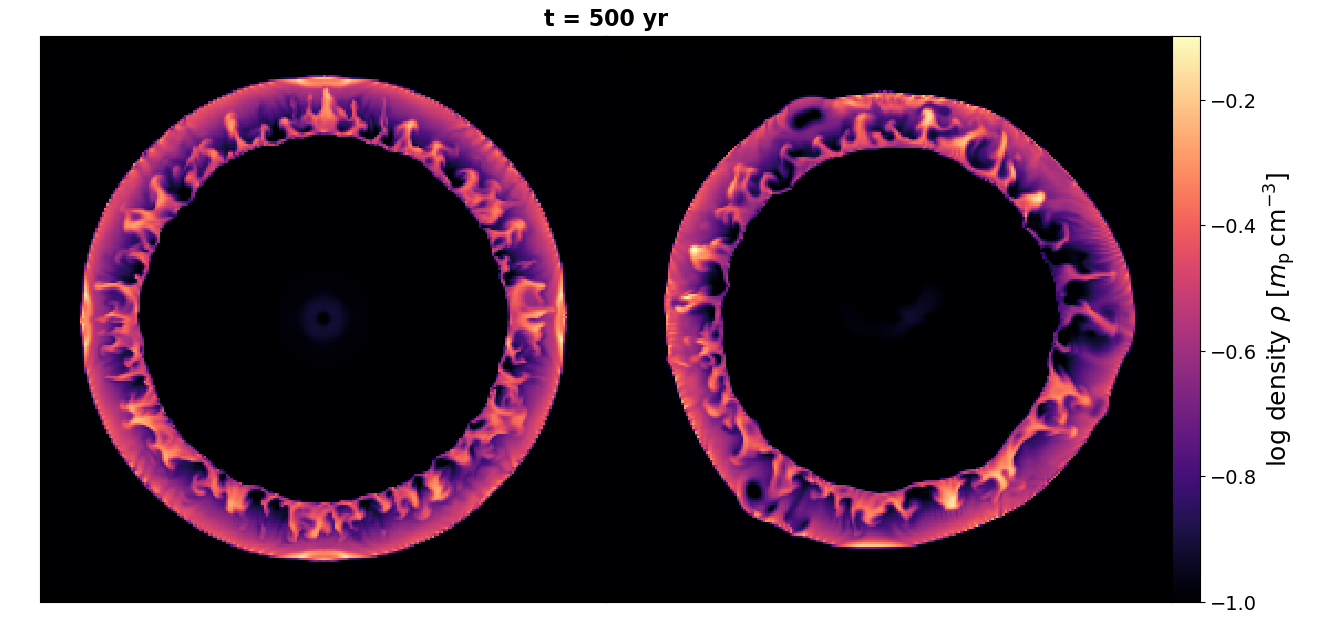}
\caption{\label{fig:SNR-1D3D-time}
Slices of the mass density at $t=1$~yr (top), 100~yr (middle), and 500~yr (bottom). 
The left side shows the case of spherically symmetric ejecta (effectively 1D initial conditions), while the right side shows the case of asymmetric ejecta (fully 3D initial conditions). An animation in time from 1~yr to 500~yr by steps of 1~yr is available online. Note that the colour scale is logarithmic and that its upper value is adjusted over time so that all frames have similar contrast (the density in the inner ejecta decreases by several orders of magnitude over this period). The simulation is done in a co-expanding grid, the linear size of the physical box increases by a factor of about 150 over this period. The box size may be slightly different at a given age for the two cases, it is of the order of 0.085~pc at 1~yr, 5~pc at 100~yr, and 13~pc at 500~yr.
}
\end{figure}

We use the output from the SN simulations described in the previous section as initial conditions for the SNR simulations. 
We systematically perform two kinds of simulations, one using the angle-averaged radial profile shown in Figure~\ref{fig:rho-1D}, mapped to the 3D grid in all directions, and one using the actual spatial distribution shown in Figure~\ref{fig:rho-2D}. Thus the initial conditions are effectively 1D (radial only) in the former case (we will refer to this as the ``1Di'' case), while fully 3D in the latter case (we will refer to this as the ``3Di'' case). The CD, the interface between the ejecta and the ISM, is unstable and develops Rayleigh-Taylor (RT) fingers even when starting from ``clean'' profiles (see Figure~\ref{fig:SNR-1D3D-time}).\footnote{On our Cartesian grid, the RTI is naturally seeded by numerical noise stemming from discretization errors.} 
The location of the ejecta in the simulation box is known at any time thanks to a passive scalar $f$, set to 1 inside the ejecta at the start time and advected with the fluid.

To run the SNR simulation we need to specify the ambient medium (in the SN simulation, the explosion happens in a quasi-vacuum). For simplicity it is assumed  to be uniform, with hydrogen density $n_H = 0.1$ cm$^{-3}$, equivalent to a total mass density $\rho_\mathrm{ISM} = 1.4\,n_H m_p$ assuming standard ISM composition. This choice is motivated so that the values of the physical quantities are representative of Tycho's SNR \citep{Williams2013}, which is one of the most promising targets for this study amongst young Galactic SNRs (age $\sim 450$~yr, radius $\sim 3-6$~pc depending on actual distance). 
The characteristic scales of the problem are controlled by the three quantities $E_\mathrm{SN}$, $M_\mathrm{ej}$, and $\rho_\mathrm{ISM}$ (see e.g. \cite{Truelove1999}), we define them here in the same way as in \cite{Warren2013}:
\begin{eqnarray}
r_\mathrm{ch} &=& \left(\frac{3 E_\mathrm{SN}}{4 \pi \rho_\mathrm{ISM}}\right)^{1/3} \approx 4.6\;\mathrm{pc}\;, \label{eq:r_ch} \\
u_\mathrm{ch} &=& \left(\frac{2E_\mathrm{SN}}{M_\mathrm{ej}}\right)^{1/2} \approx 10,000\;\mathrm{km\,s}^{-1}\;, \label{eq:u_ch} \\
t_\mathrm{ch} &=& \frac{r_\mathrm{ch}}{u_\mathrm{ch}} \approx 450\;\mathrm{yr}\;. \label{eq:t_ch}
\end{eqnarray}
Since no micro-physics is included, the simulations presented in this paper may be re-scaled to other sets of initial conditions using relations (\ref{eq:r_ch}), (\ref{eq:u_ch}), (\ref{eq:t_ch}). 

\paragraph{Time evolution.}

Assuming self-similarity of the early evolution, we start the simulations at $t_0=1$~day, after simply re-scaling the SN data from 100~s to the SNR start time. Self-similarity implies that radial distance scales as time~$t$, while mass density gets diluted as~$t^{-3}$. We also performed simulations starting at $t_0=1$~year, there are no significant differences visible on the SNR structure (and thus on the maps and spectra introduced in the next section) after a few years of evolution, including the numbers and sizes of the RT fingers. This shows that self-similarity is indeed a safe assumption for the very young SNR. We present the results up to 500~yr of evolution, similar to the age of Tycho's  SNR. 
We emphasize that our purpose here is to study the SNR phase (from a realistic SN model), not the SN phase per~se. In particular we are not concerned with the modeling of the light curve, which requires complex radiative transfer simulations to describe the first days/weeks after the explosion (see \cite{Sim2013} for the N100 model). However we checked the possible effect of heating from radioactive decay, see the discussion in Section~\ref{sec:disc-add} and more details in Appendix~\ref{sec:decay}. Radiative cooling is not included in the simulation, since it is negligible for this phase of the SNR dynamical evolution.

One distinctive technical aspect of the simulations by \cite{Ferrand2010}, re-used and extended here, is that we are working in an expanding grid. The number of cells is fixed, but their physical size increases, so that the SNR keeps about the same size relative to the grid.
Factoring out the global evolution of the remnant allows us to maintain the same relative resolution in the ejecta and shocked region. From 1~day to 500~years, the SNR extends by a factor of about 50,000. The principle of the comoving transformation with equations is given in Appendix~\ref{sec:comoving}. It affects all variables, hydro quantities as well as space and time. The key quantity is the scale factor $a(t)$, which describes the physical grid size as a function of time. In previous works, we considered the ideal case where the scale factor can be written as a power-law $a(t) \propto t^\lambda$ with constant parameter~$\lambda$. New in this work, we simulate the SNR as it smoothly transitions from the free-expansion phase into the ejecta-dominated phase, up to the time when the swept-up CSM becomes important (the Sedov-Taylor time for our model SNR happens to be about 500~yr). We are therefore using a variable expansion law. Then no simple analytical relation exists between comoving time and physical time, we numerically compute the comoving time as a function of physical time given the observed scale factor $a(t)$. See details in Appendix~\ref{sec:comoving}.

\subsection{Analysis of the SNR morphology}
\label{sec:waves}

The main focus of our work is on the (3D) morphology of the SNR. 
Young SNRs are best observed in the X-ray domain, where one can image the emission from the hot plasma \citep{Vink2012}. We therefore focus our analysis on the shocked region, bounded by the reverse and forward shocks.
While most SNR studies assume spherical symmetry, we want to know how the shocks develop from a more realistic ejecta distribution. In between the shocks, the ejecta are delimited by the CD, where the RTI generates turbulence. As 3D simulations of SN explosions have become available, a question arises, that to our knowledge has not been addressed yet, of how the RTI of the SNR phase develops on top of the three-dimensional structures resulting from the SN phase. 

To investigate the signature of the N100 SN model, we thus choose to analyze over time the morphology of the three wave fronts: forward shock (FS), contact discontinuity (CD), reverse shock (RS). The CD is located using the ejecta fraction, while the shocks (RS and FS) are detected as pressure jumps. Since we are working in a Cartesian grid, we extract a three-dimensional mesh made of quadrilateral cells. Our simple detection scheme ensures that we extract a contiguous mesh for each wave surface. Implementation details are given in Appendix~\ref{sec:waves-analysis}. 

The location of each wave (radial distance from the SN center) is then projected on the surface of a sphere for visualization. For the tessellation of the sphere we are using the efficient HEALPix scheme \citep{Gorski2005}. 
The maps in the left panels of Figures~\ref{fig:healpix_CD}, \ref{fig:healpix_RS}, and \ref{fig:healpix_FS} show the normalized wave radius $R=(r-\langle r\rangle)/\langle r\rangle$, where $\langle r\rangle$ is the average over all directions at a given time. (It is this average value, for the CD, that defines the global expansion law of the grid.) 
We use the Mollweide projection of the sphere, which is an equal area projection.
We checked that the maps presented here are converged w.r.t. the projection of the 3D meshes.

Finally, to quantify the asymmetries observed, we expand the function $R(\theta,\phi)$ in spherical harmonics. The spectra in the right panels of Figures~\ref{fig:healpix_CD}, \ref{fig:healpix_RS}, and \ref{fig:healpix_FS} decompose the wave surface as a function of the angular scale of the fluctuations of~$R$.\footnote{We chose to analyze three surfaces of particular significance. In principle one could also analyze the 3D density field as a whole, by performing a spherical Fourier-Bessel decomposition, with spherical Bessel functions $j_{\ell}(kr)$ describing the radial part and spherical harmonics $Y_{\ell}^m(\theta,\phi)$ describing the angular part \citep{Leistedt2012}.} At angular wavenumber~$\ell$, the typical angular scale probed is $\pi/\ell$. The power $C_{\ell}$ plotted is normalized in such a way that each grayed bin is the contribution of wavenumber~$\ell$ to the total variance of~$R$ over the sphere (see Appendix~\ref{sec:waves-analysis} for details).

\section{Results}
\label{sec:results}

\def\healpixwidth{1.}

\begin{figure}[t!]
\gridline{\fig{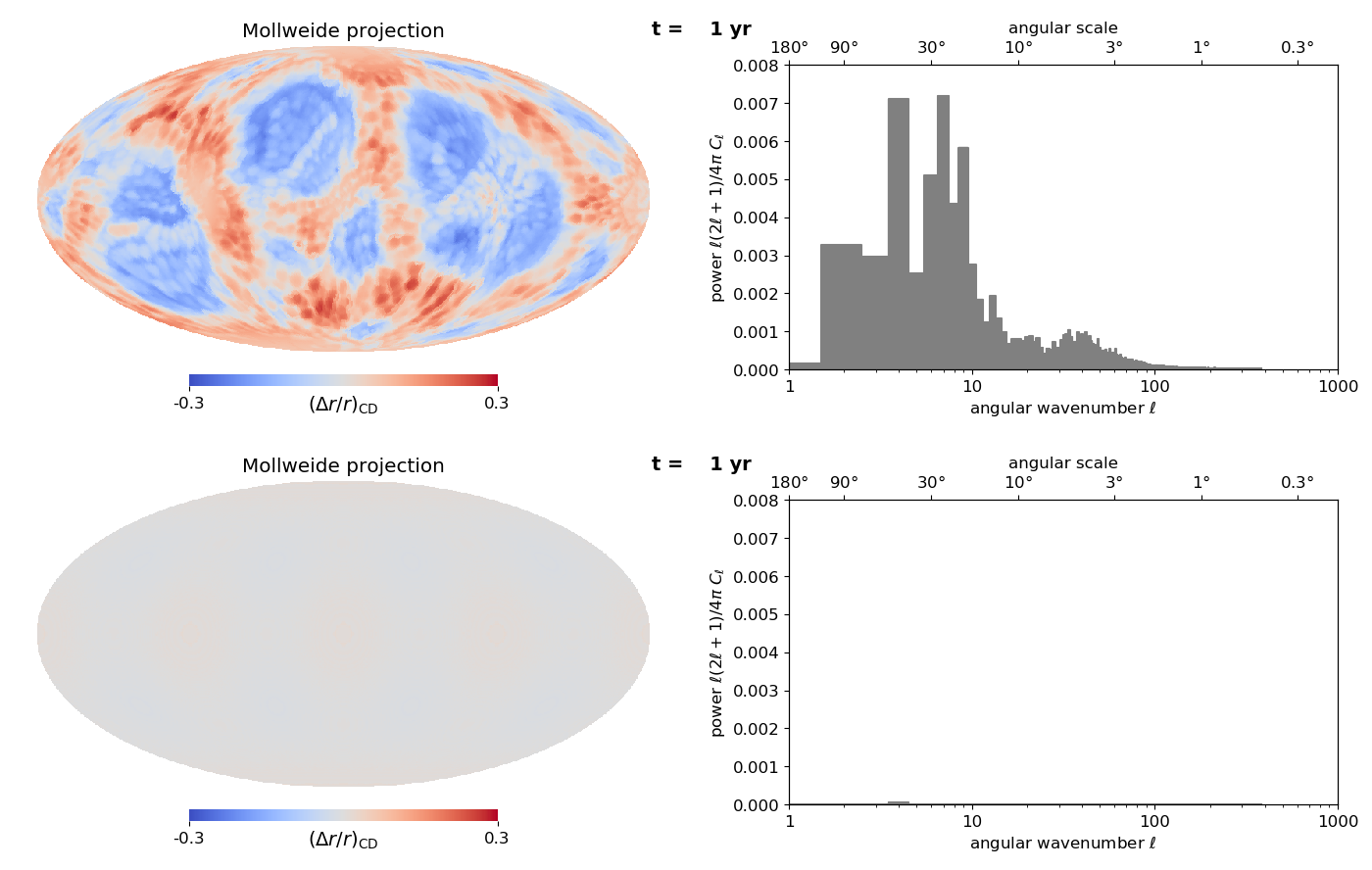}{\healpixwidth\textwidth}{(a) Contact discontinuity at 1~yr}}
\caption{Morphology of the contact discontinuity. Maps on the left are spherical projections of the radial variations of the location of the wave. Spectra on the right result from an expansion in spherical harmonics of these variations. As before, two cases are compared: spherically symmetric ejecta (effectively 1D initial conditions, at the bottom) versus asymmetric ejecta (fully 3D initial conditions, at the top). 
Three times are shown: 1~yr (a), 100~yr (b), and 500~yr (c). 
An animation in time from 1~yr to 500~yr by steps of 1~yr is available online.}
\label{fig:healpix_CD}
\end{figure}
\renewcommand{\thefigure}{\arabic{figure} (continued)}
\addtocounter{figure}{-1}
\begin{figure}[t!]
\gridline{\fig{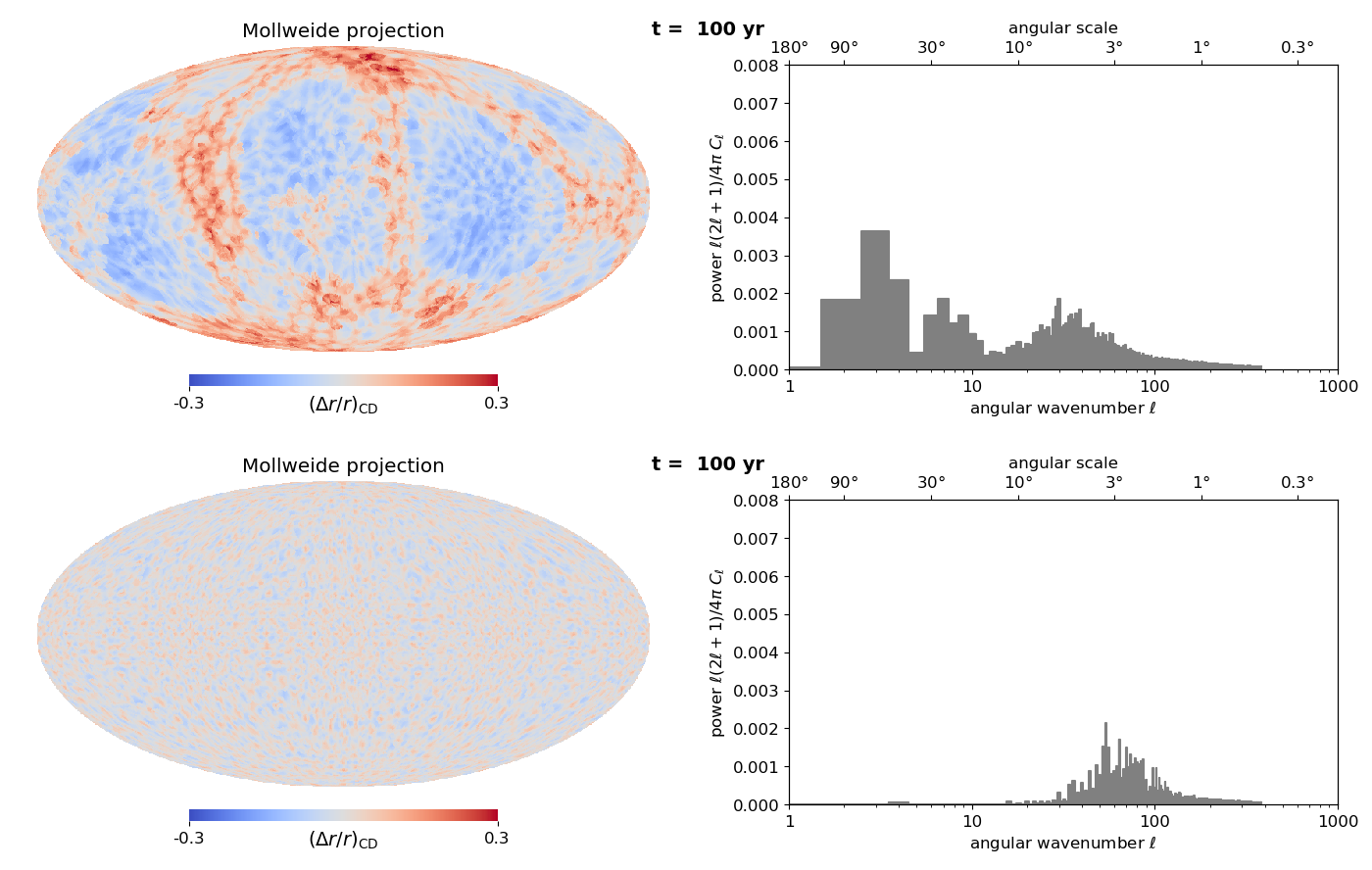}{\healpixwidth\textwidth}{(b) Contact discontinuity at 100~yr}}
\vspace{3mm}
\gridline{\fig{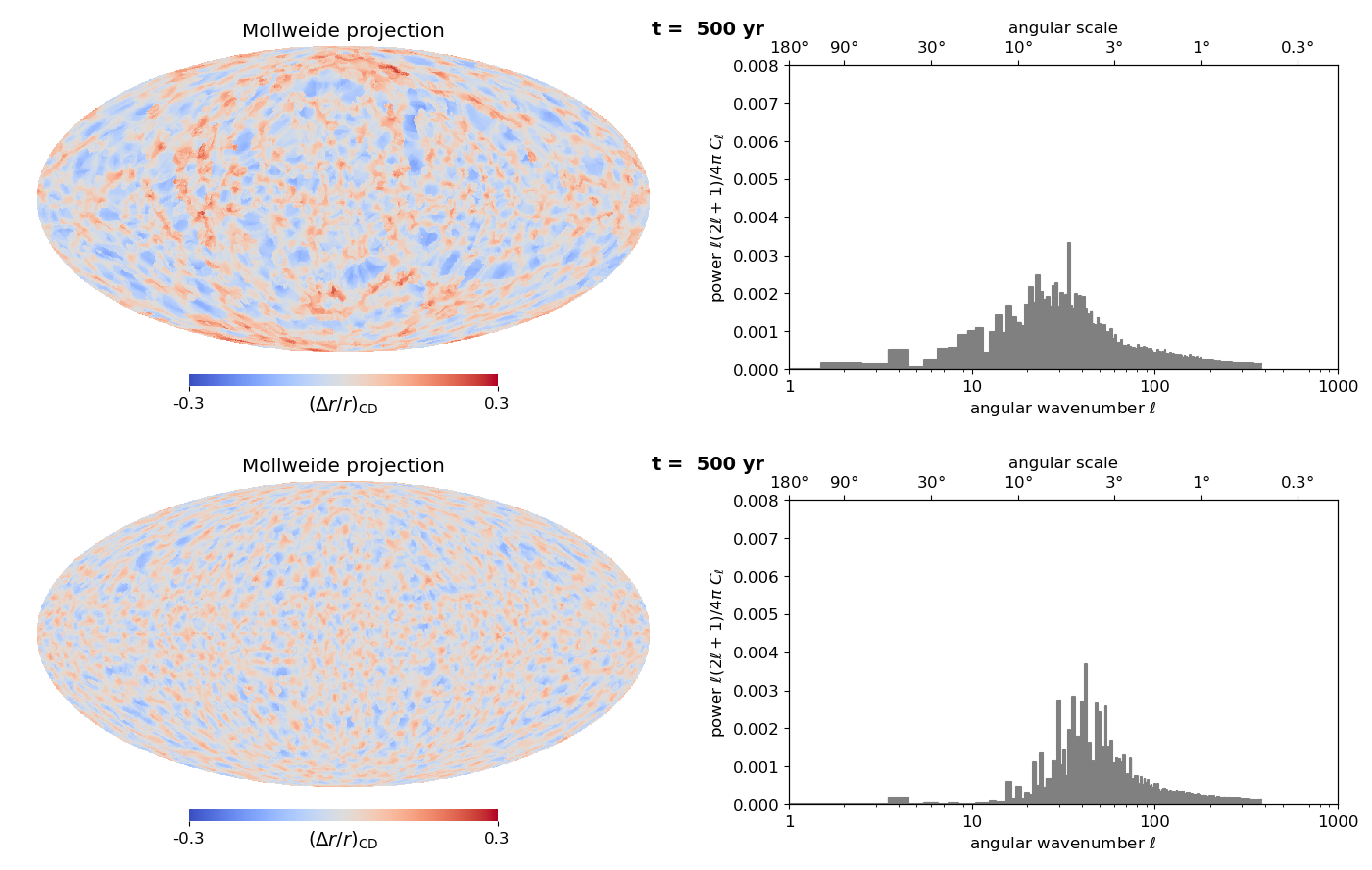}{\healpixwidth\textwidth}{(c) Contact discontinuity at 500~yr}}
\caption{}
\end{figure}
\renewcommand{\thefigure}{\arabic{figure}}

\begin{figure}[t!]
\gridline{\fig{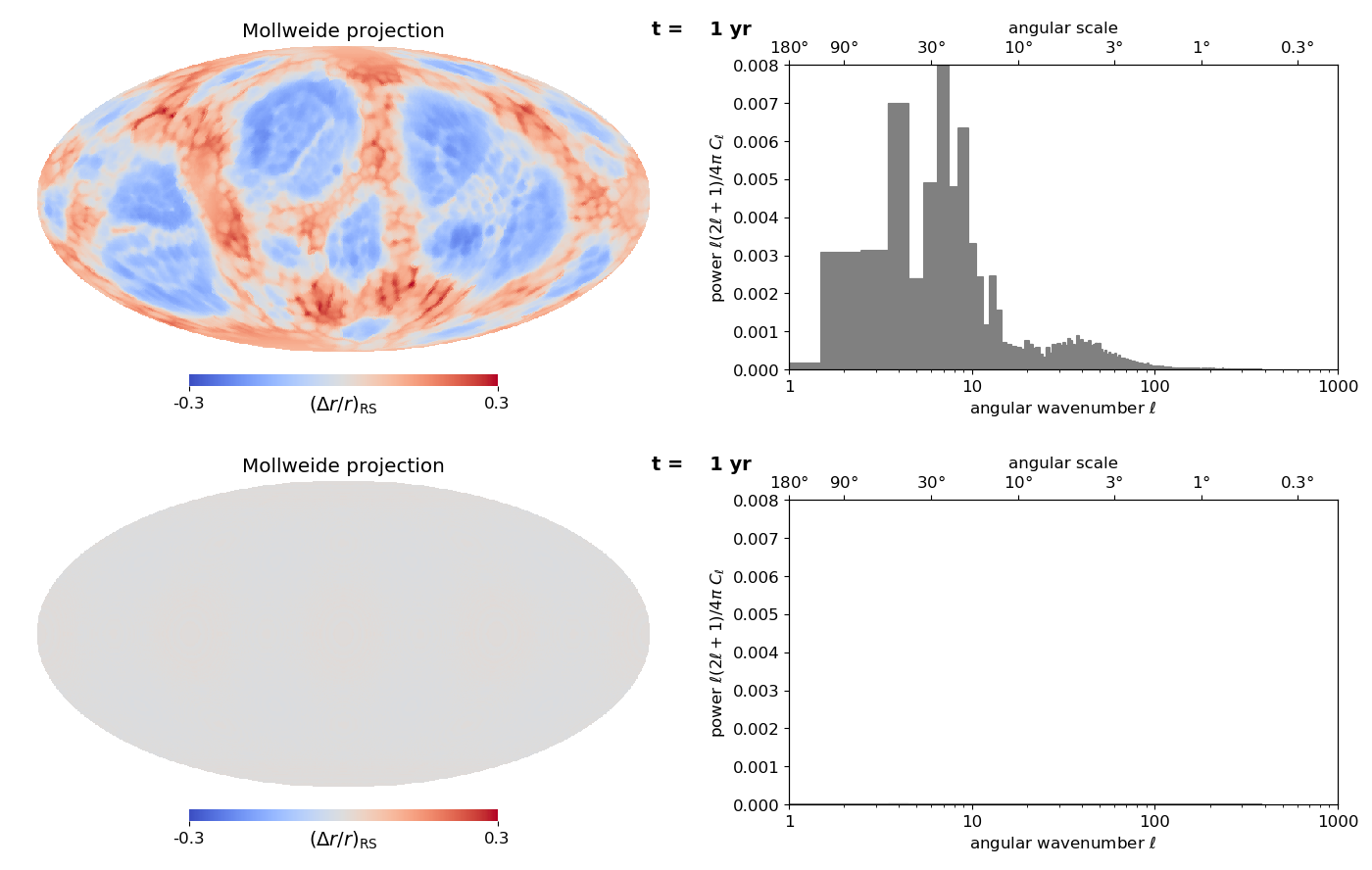}{\healpixwidth\textwidth}{(a) Reverse shock at 1~yr}}
\caption{Morphology of the reverse shock. Projected maps and angular spectra are the same as in Figure~\ref{fig:healpix_CD}. 
Three times are shown: 1~yr (a), 100~yr (b), and 500~yr (c).
An animation in time from 1~yr to 500~yr by steps of 1~yr is available online.}
\label{fig:healpix_RS}
\end{figure}
\renewcommand{\thefigure}{\arabic{figure} (continued)}
\addtocounter{figure}{-1}
\begin{figure}[t!]
\gridline{\fig{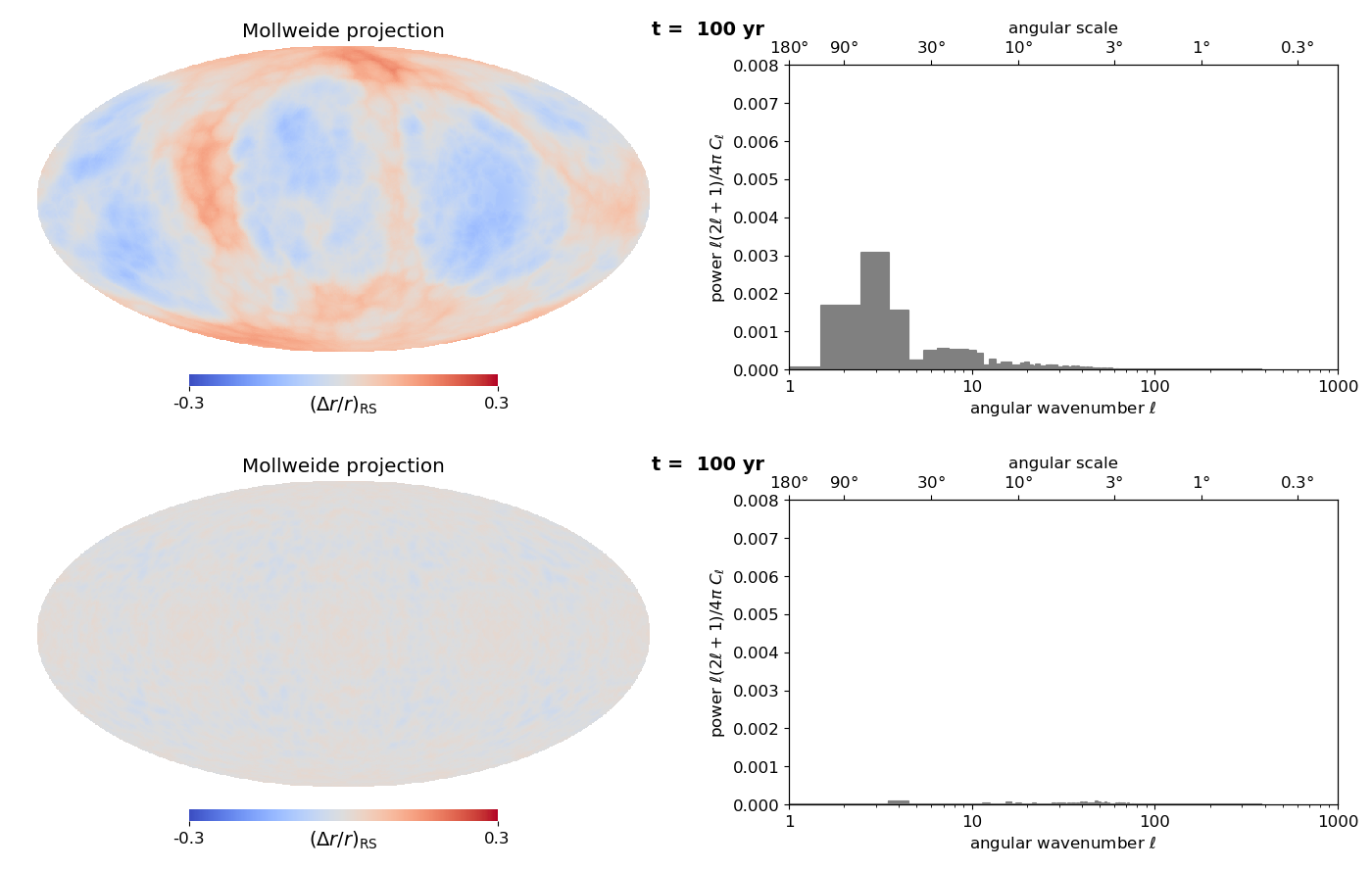}{\healpixwidth\textwidth}{(b) Reverse shock at 100~yr}}
\vspace{3mm}
\gridline{\fig{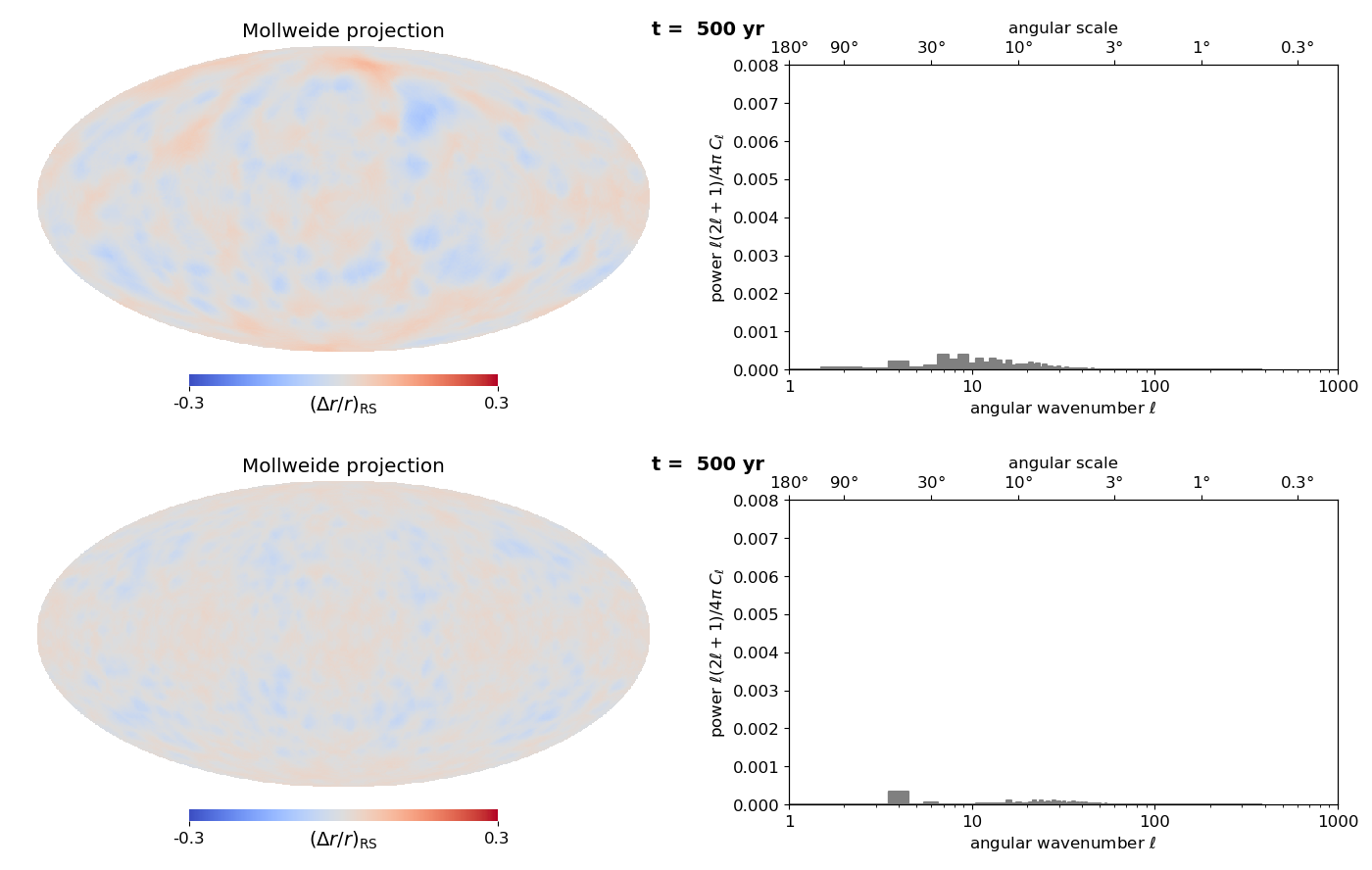}{\healpixwidth\textwidth}{(c) Reverse shock at 500~yr}}
\caption{}
\end{figure}
\renewcommand{\thefigure}{\arabic{figure}}

\begin{figure}[t!]
\gridline{\fig{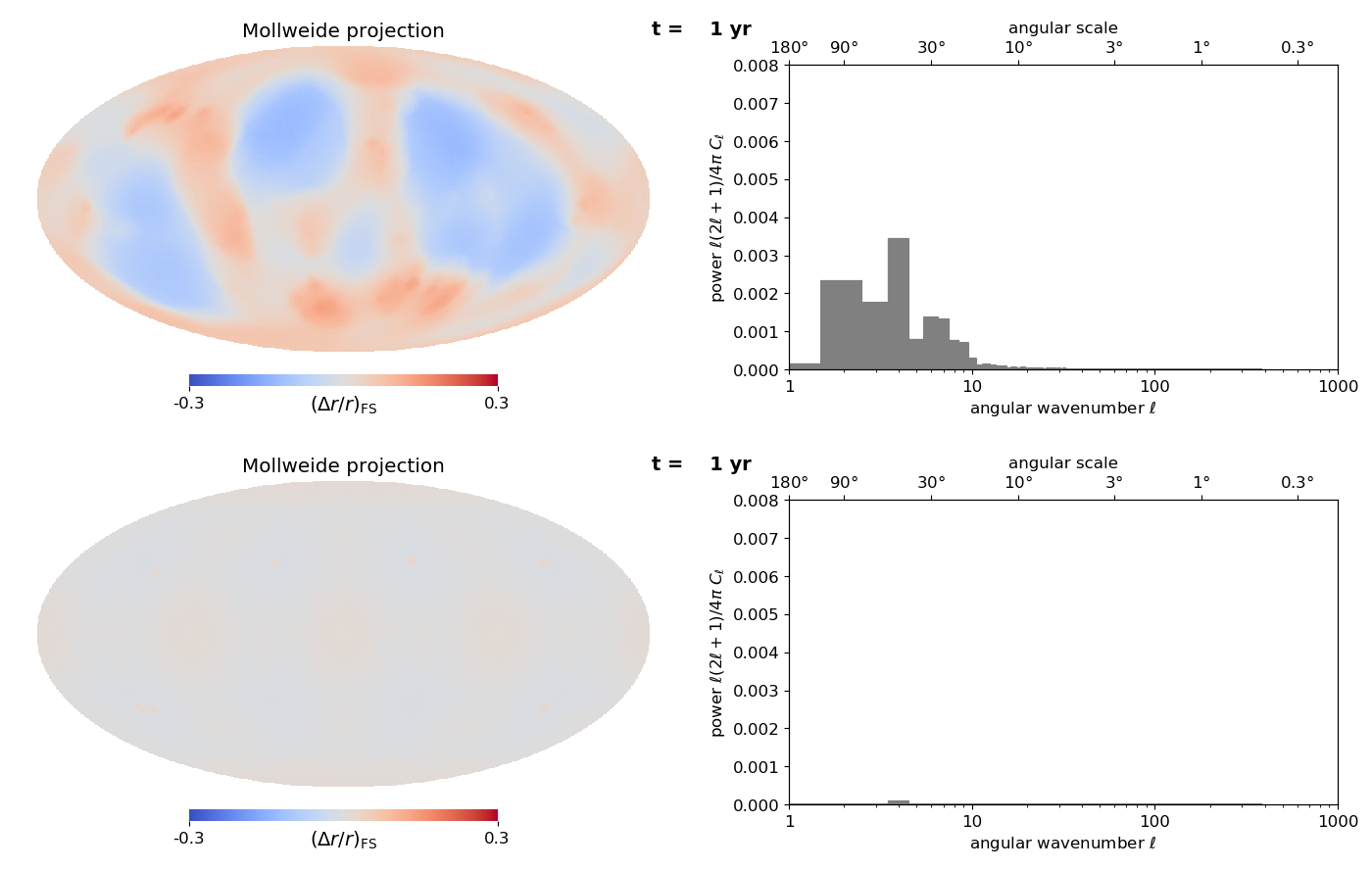}{\healpixwidth\textwidth}{(a) Forward shock at 1~yr}}
\caption{Morphology of the forward shock. Projected maps and angular spectra are the same as in Figure~\ref{fig:healpix_CD}.
Three times are shown: 1~yr (a), 100~yr (b), and 500~yr (c).
An animation in time from 1~yr to 500~yr by steps of 1~yr is available online.}
\label{fig:healpix_FS}
\end{figure}
\renewcommand{\thefigure}{\arabic{figure} (continued)}
\addtocounter{figure}{-1}
\begin{figure}[t!]
\gridline{\fig{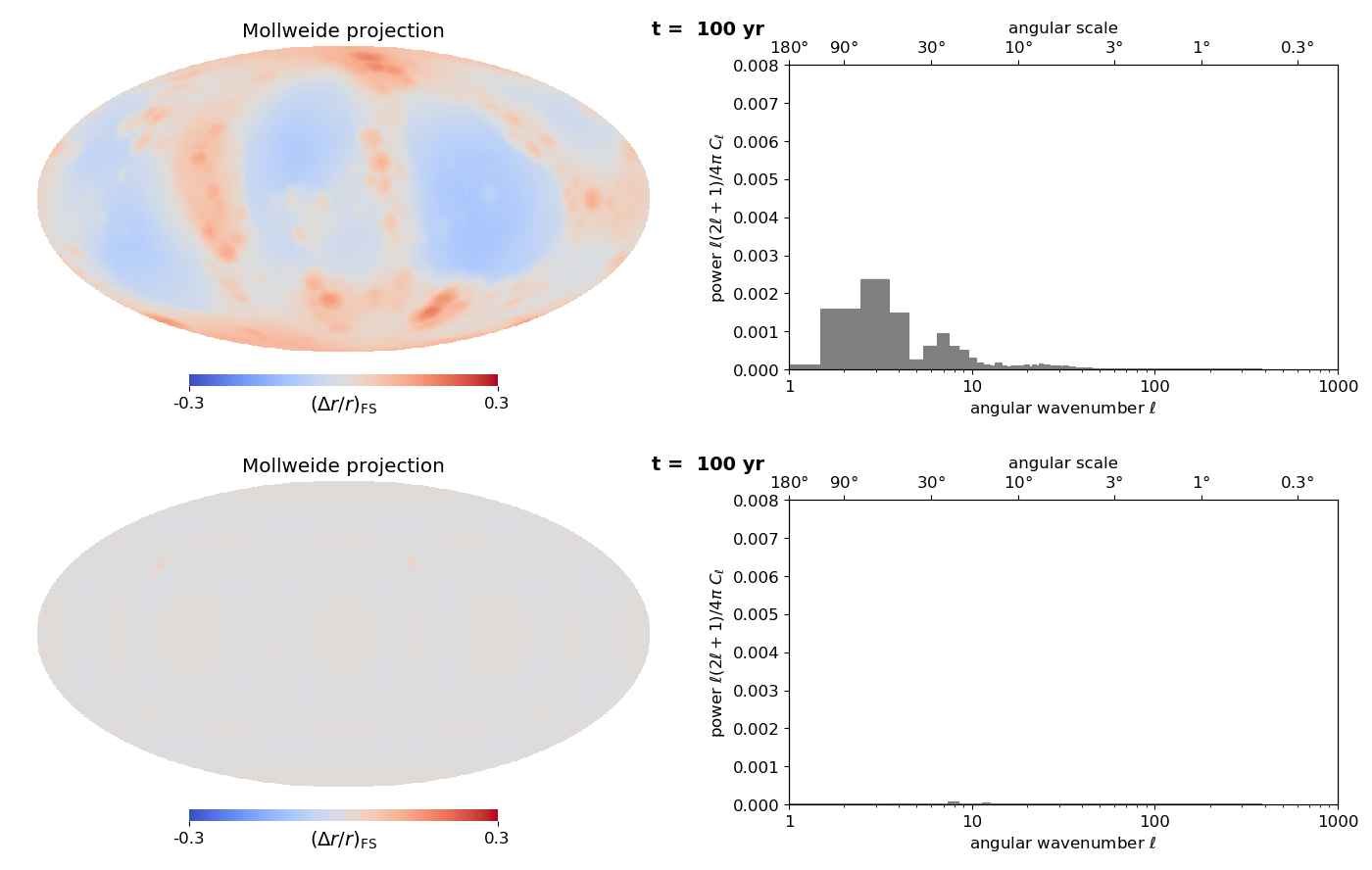}{\healpixwidth\textwidth}{(b) Forward shock at 100~yr}}
\vspace{3mm}
\gridline{\fig{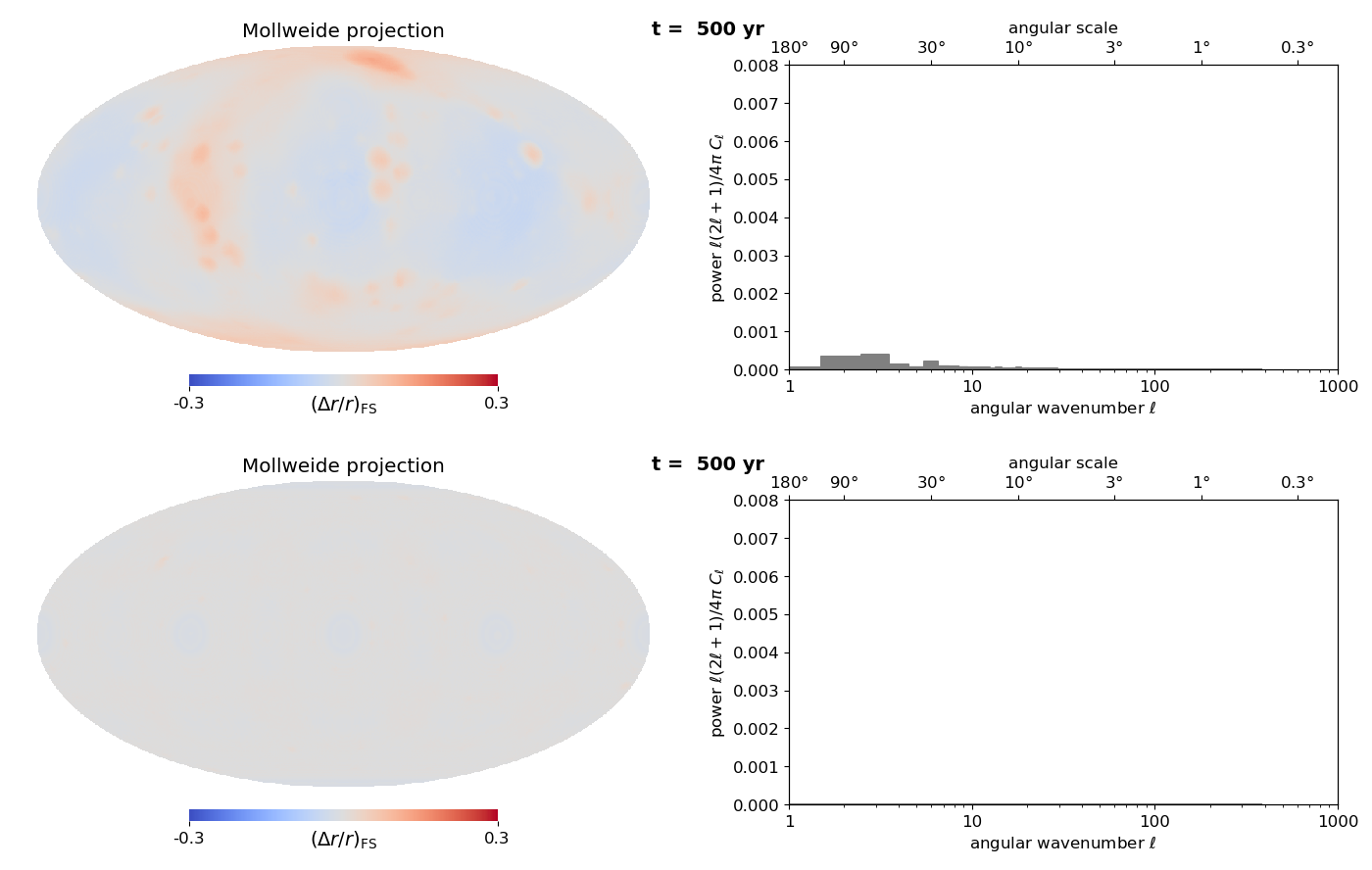}{\healpixwidth\textwidth}{(c) Forward shock at 500~yr}}
\caption{}
\end{figure}
\renewcommand{\thefigure}{\arabic{figure}}

We are now equipped to examine the time evolution of the SNR resulting from the N100 SN model.

\paragraph{Density maps} In order to show the inner structure of the remnant, we present slices of the mass density field in Figure~\ref{fig:SNR-1D3D-time}. A movie from 1~yr to 500~yr is available online, we show here three representative snapshots at 1~yr, 100~yr, and 500~yr.\footnote{A volume rendering of the 3D data cube, evolving in time, can also be explored in the VR demo previously mentioned, as explained in \cite{Ferrand2018}.} At 1~yr (top snapshot) the ejecta distribution is still very similar to the initial distribution assumed at 1~day. The shocks are present on this plot, although barely visible even on a logarithmic scale. The FS actually appears on the first few time steps, quickly followed by the RS. At 100~yr (middle snapshot) the shell of shocked material, bounded by the RS and the FS, becomes more apparent; at 500~yr (bottom snapshot) it is all that can be seen on the map. From about 10~yr the RTI becomes visible by eye, producing the characteristic pattern of holes and fingers along the CD.

On all the snapshots we compare the two types of initial conditions: effectively 1D on the left (1Di case, depending on~$r$) versus fully 3D on the right (3Di case, depending on $r$, $\theta$, $\phi$). Maps on the left are representative of previous works on SNRs (e.g. \cite{Blondin2001,Ferrand2010,Warren2013}). Maps on the right are more realistic, being obtained from an actual SN simulation. And encouragingly, they do look closer to the morphology of a young SNR like Tycho (see \cite{Warren2005} for the X-ray images, more on this in the next section).
The 1Di case of spherically symmetric initial conditions is used as a reference. It shows the RTI generated purely from the SNR phase, while the 3Di case shows the RTI growing on top of ejecta with an already complex morphology. This is particularly obvious in the first hundred years of evolution, less so by the final time of 500~yr. One can see how the details of the SN explosion, that is the initial shape of the ejecta, are progressively forgotten over time. Still, the SN imprint is clearly visible at 100~yr, and not fully erased at 500~yr. 
We also note that, when ejecta have asymmetries from the start, it appears easier for the RT fingers to get into contact with the FS, a phenomenon that has been observed in young SNRs but has proven difficult to reproduce without introducing pre-existing structures \citep{Jun1996} or increasing the compressibility of the fluid \citep{Blondin2001}. This is in line with the findings of \cite{Orlando2012} on the role of ejecta clumping. 

\paragraph{Projected maps and power spectra}
The relative variations of the position of the CD, RS, and FS are shown in Figures~\ref{fig:healpix_CD}, \ref{fig:healpix_RS}, and \ref{fig:healpix_FS} respectively, at the same times as in Figure~\ref{fig:SNR-1D3D-time} (movies from 1~yr to 500~yr are available online). Maps on the left sides allow for visual inspection of (the 3D surface of) the waves, while spectra on the right sides allow for quantification of the asymmetries.
Note that the scales used for the maps and for the spectra are the same on all plots, across the different kinds of waves (CD, RS, FS), initial conditions (1Di vs. 3Di), and times (from 1~yr to 500~yr), to facilitate comparison.

For the shocks, the 1Di case provides a measure of the numerical precision of our code, especially the FS where nothing is expected to happen. We see that the only measurable anomaly is a small $\ell=4$ mode from the cubic geometry of the hydro grid. For the more realistic 3Di case, we see how the shocks first follow the contours of the ejecta, then homogenize in radius thanks to the high pressure in the shocked material. 
At late times the RS shows slightly rising power at $\ell\geq10$. At 500~yr the RS is still very close from the CD (see Figure~\ref{fig:SNR-1D3D-time}), and their surfaces are well correlated: we mostly see the feet of the RT fingers. 
At 500~yr the FS still shows some protrusions, at scales $\ell<10$. Again comparing with the CD surface, we see that these correspond to the tips of elongated RT fingers, growing along large ridges of ejecta that were present from the beginning. This confirms that the clumpiness of the ejecta enhances their interaction with the FS. We note that, in such a young SNR, the CD as well as the shock fronts probe the outer layers of the ejecta, and thus mostly the effect of the detonation fronts that ended the SN explosion. 

For the CD, we observe the same overall regularizing of the initial morphology, plus the RTI growing on top of it. The 1Di case (bottom half of the figures) shows the RTI from the SNR phase, while the 3Di case (top half of the figures) shows the RTI from the SN+SNR phases, comparison between the two allows us to separate the contribution from the SN phase.
 The SN power is present at small~$\ell$ (large angular scales), simply going down in time, while the RTI power is present at larger~$\ell$ (small angular scales), going up as well as shifting to smaller~$\ell$ in time. This is the expected behavior: the growth rate of the RT instability (in the initial linear regime) is known to scale as the square root of the wavenumber of a perturbation, so it first appears at the smallest scales resolved and grows to larger scales over time (initially by finger growth, then through finger mergers). On the first time frame, we see that the instability had actually started to develop by the end of the SN simulation. 

\begin{figure}[ht!]
\centering
\includegraphics[width=0.85\textwidth]{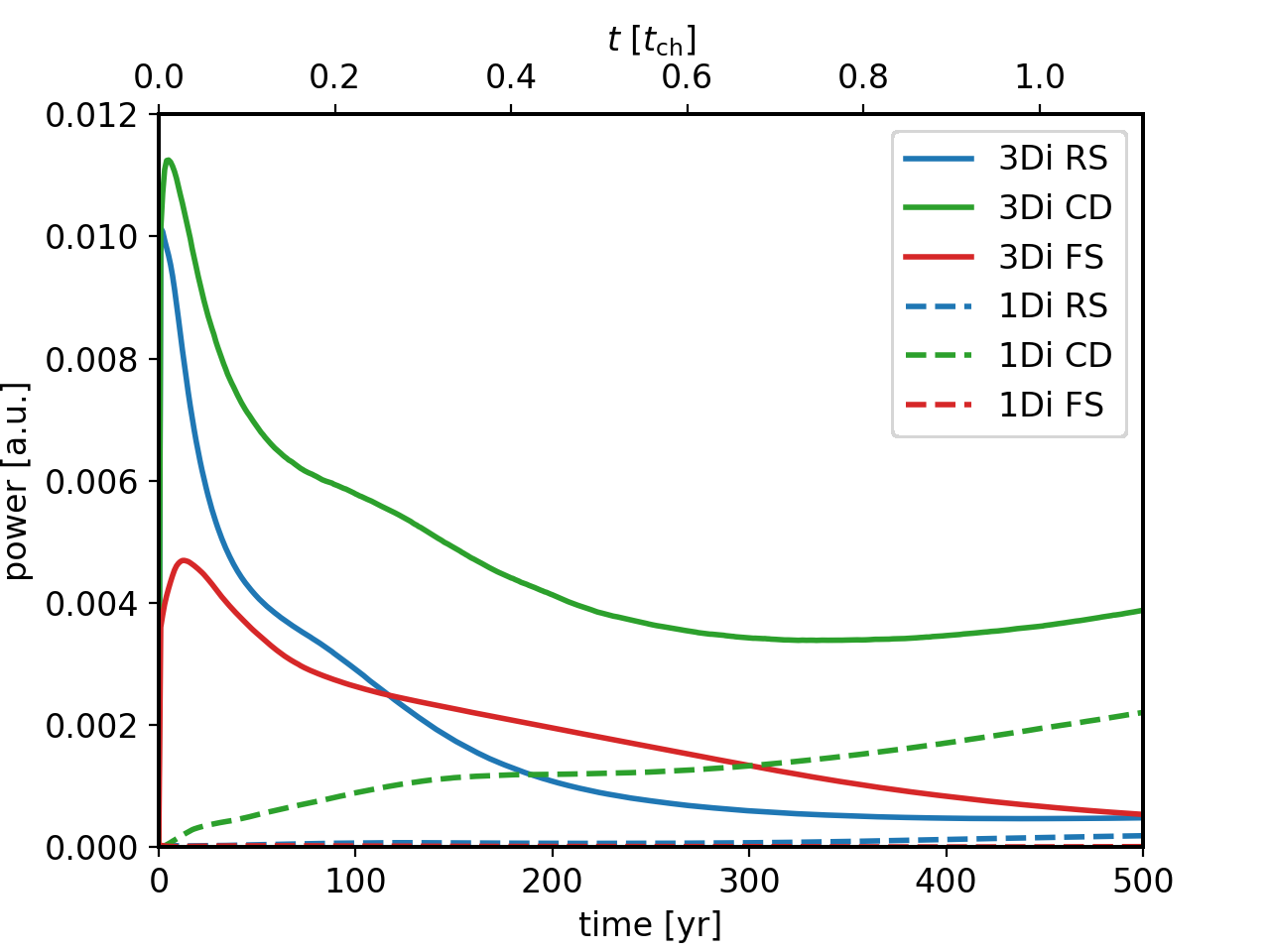}
\includegraphics[width=0.85\textwidth]{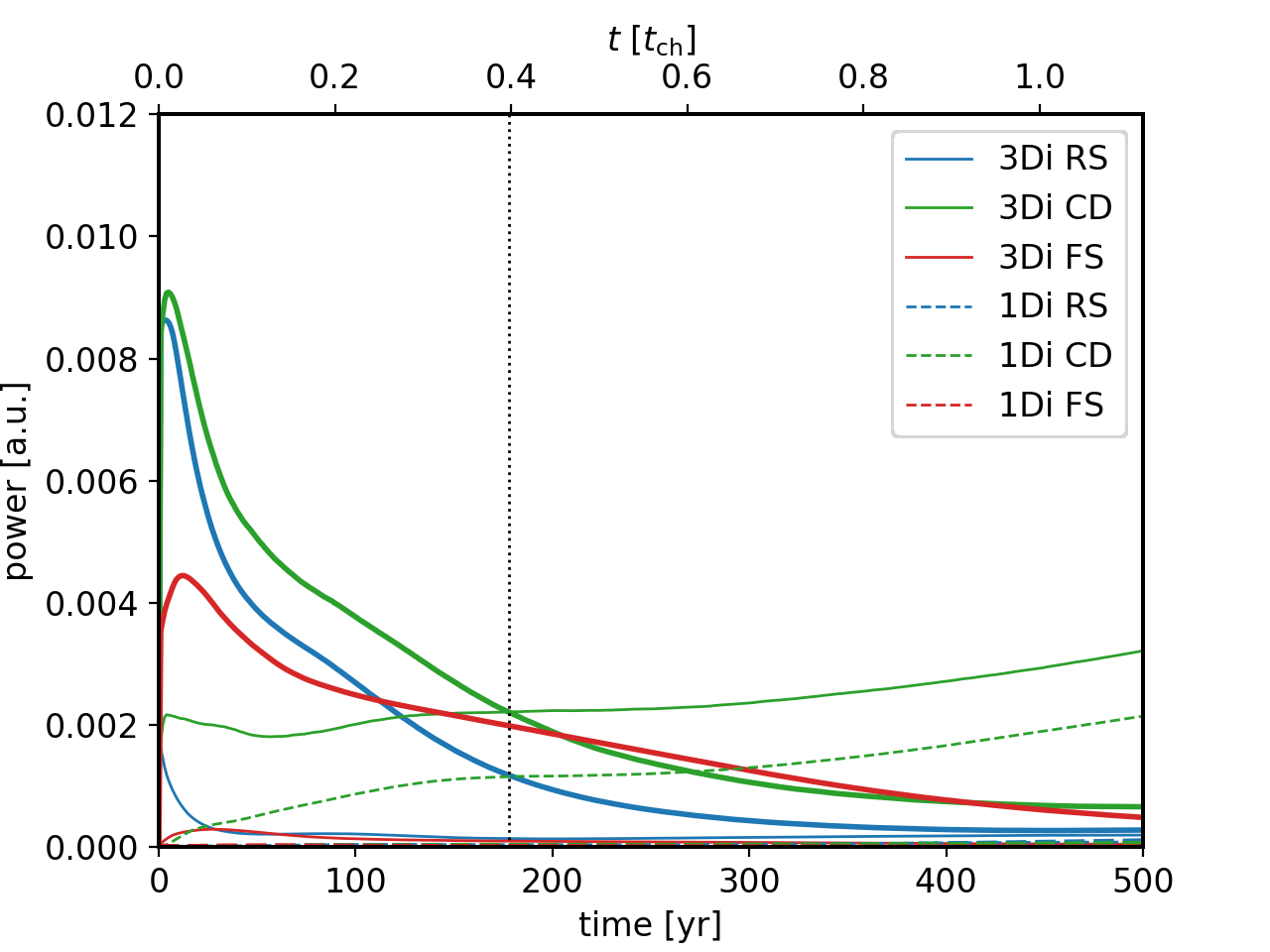}
\caption{\label{fig:power_time}
Evolution of the power as function of time, for the three waves: FS in red, CD in green, RS in blue. Time is indicated in years and in characteristic timescale $t_\mathrm{ch}$ defined by Eq.~(\ref{eq:t_ch}). Two cases are compared: spherically symmetric ejecta (1Di, dashed curves) versus asymmetric ejecta (3Di, solid lines). 
Top: the total power is plotted, integrating all angular scales (sum of the gray area on the corresponding power spectra plots), bottom: the power is plotted separately at large scales ($\ell<10$, thick lines) and at small scales ($\ell>10$, thin lines). The vertical dotted line indicates the time at which the two curves intersect for the CD from 3D initial conditions. }
\end{figure}

\paragraph{Total power as a function of time}
The time evolution of the total power in the fluctuations of the wave positions is shown in Figure~\ref{fig:power_time} (top panel, this corresponds to the gray area on the power spectra). 
Looking first at the 1Di case (dashed lines), the curve for the FS is a measure of the numerical precision, the curve for the CD is purely due to the RTI, while the curve for the RS shows a faint imprint of the RTI. 
Looking now at the 3Di case (solid lines), we see the trends discussed with maps and spectra. For the shocks, the power is continuously decaying. For the CD, it also decays until about 300~yr, when it starts to increase. It then follows the same temporal evolution as in the 1Di case, showing that power is now dominated by the RTI of the SNR phase. 

As shown in the previous paragraphs, from the comparison of the power spectra between the 1Di and 3Di cases one can separate the ranges of $\ell$ for which the SN phase and the SNR phase dominate. In the bottom panel of Figure~\ref{fig:power_time} we thus plot the integrated power for $\ell<10$ (large scales, thick lines) and $\ell>10$ (small scales, thin lines). Let us concentrate on the evolution of the CD (green lines). We see that the power at small scales in the 1Di case (thin dashed line) is the same as the total power in this case (dashed line of the top plot). And the power at small scales in the 3Di case (thin solid line) follows the same time evolution as in the 1Di case (thin dashed line). This validates our choice of $\ell=10$ to isolate the RTI. Then, the extra power at large scales for the 3Di case shows the contribution of the SN (thick solid line). This component is only decaying over time. Curves for the CD power at $\ell<10$ and at $\ell>10$ cross at $t=178$~yr = $0.4\,t_\mathrm{ch}$ (thick vs. thin solid green lines), which is our estimate of the transition time between an SN-dominated morphology and an SNR-dominated morphology for the N100 model. 

\section{Discussion}
\label{sec:discussion}

Our analysis has shown that the SN explosion can have a measurable impact on the morphology of the subsequent SNR. We now discuss some numerical and physical effects that may impact our results, and their applicability to observations.

\subsection{Numerical robustness}
\label{sec:disc-num}

We first assess the robustness of our findings with respect to the particular simulation setup we have used.

\paragraph{Effect of numerical resolution.}
Our simulations and analysis are performed at a resolution appropriate for the data cubes that we obtained from the SN simulation. 
To check the effect of spatial resolution, we varied the resolution of the Cartesian grid by a factor of two lower and higher (along each dimension). Using the effectively 1D initial conditions (1Di model), as expected RT fingers appear at increasing $\ell$ as the spatial resolution increases (they will always appear at the smallest scale resolved), however after a few tens of years the power spectrum looks similar, and most importantly the smallest $\ell$ reached is the same as a function of time (meaning that the start time is early enough to properly catch the RTI at these scales); in particular it never reaches below $\ell=11$ (except for the small and known to be artificial $\ell=4$ mode). Using the fully 3D initial conditions (3Di model), differences between the spectra are small when doubling or halving the spatial resolution, and all the previous discussion still holds. So we are confident that the general trends observed on the maps and spectra presented here are revealing the physics of the SN to SNR connection.

We also comment on the use of adaptive mesh refinement (AMR), which is offered by the RAMSES code in the form of octree refinement. To increase the resolution in the most interesting regions, near the shocks and the CD, one can set refinement rules based on gradients of respectively the pressure and the ejecta fraction, as was done for the \cite{Ferrand2010} simulations. This technique was not used in the simulations presented in this paper, since it was found to introduce some noise in the spectra, for a limited extra gain in computing time. However the diagnostics we are discussing, in terms of overall distribution of scales as a function of time, were not significantly affected.

\paragraph{Effect of grid geometry.}
Both the SN and the SNR simulations presented here were performed in Cartesian geometry. 
In such a geometry, slightly different behaviors may be observed when shock waves are propagating along the principal axes versus along a diagonal. This spurious effect is visible at the outer edge of the simulated SNR on Figure~\ref{fig:SNR-1D3D-time}, for the 1Di case at the final age. However it can be seen from the angular spectra of Figure~\ref{fig:healpix_FS} that the extracted shock surface remains spherical with sufficient numerical precision.

We now compare our results with the work by \cite{Warren2013}, who did simulations with the code VH1 in spherical geometry (using a Yin-Yang grid). They assumed spherically symmetric ejecta with an exponential profile, which can be compared with our angle-averaged initial conditions (labeled 1Di). They computed the power spectrum from a spherical harmonics expansion, after summing the ejecta density in radial columns (see their Figures~3 and~6). Although the analysis is different, it should provide similar results. 
They use a dimensionless time, in units of $t_\mathrm{ch}$ defined by Eq.~(\ref{eq:t_ch}). On their Figure~3, at times $t=0.15\;t_\mathrm{ch} \simeq 67$~yr and $t=0.75\;t_\mathrm{ch} \simeq 335$~yr, the power spectrum peaks around respectively $\ell = 60$ and $\ell = 45$, and drops below 1\% of the peak value around respectively $\ell = 20$ and $\ell = 10$, which in good agreement with our own power spectra for the 1Di case (see Figure~\ref{fig:healpix_CD}). 

\begin{figure}[t!]
\centering
\includegraphics[width=0.9\textwidth]{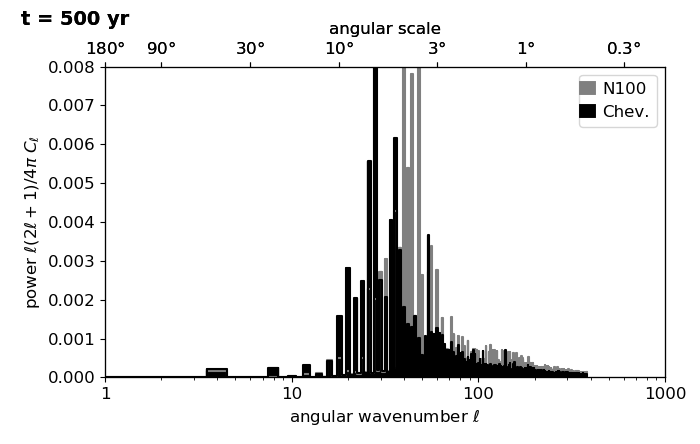}
\caption{Angular power spectrum for the contact discontinuity at the final age $t=500$~yr. Two ejecta profiles are compared (both of which are effectively 1D), that were used as initial conditions at $t=1$~yr. The gray histogram was obtained using the angularly-averaged N100 model (1Di case), the black one was obtained using the Chevalier semi-analytical profiles.}
\label{fig:healpix_CD_Chev}
\end{figure}

\paragraph{Effect of initial ejecta profile.}
As seen from Figure~\ref{fig:rho-1D} the average radial profile of our SN model is quasi-exponential. For completeness, we also tried to initialize the SNR simulation with the \cite{Chevalier1982} profile. Since this is a 1D model, we compare it to the N100 1Di case. We start the simulations at $t_0=1$~yr, since the Chevalier semi-analytical solution is not valid too early on. The width of the shock region is slightly larger at a given time, and accordingly the peak of the power spectrum is at a somewhat larger scale (lower~$\ell$), as can be seen in Figure~\ref{fig:healpix_CD_Chev}. But the minimum~$\ell$ reached by the RTI at a given age is similar, so our conclusions are unchanged, regarding the possible contributions of the SN phase versus the SNR phase. 

\subsection{Possible additional physics}
\label{sec:disc-add}

We now consider three energy reservoirs that were not included in the simulations: radioactive decay, the magnetic field, and non-thermal particles.

\paragraph{Effect of heating from radioactive decay.} 
In this work our focus is not on the SN phase per se, but on its possible impact on the SNR phase. However, one may ask about the possible role of heating from radioactive decay, which powers the optical light curve of the SN. The most important species is ${}^{56}\mathrm{Ni}$, which decays via beta decay to iron in two steps: 
${}_{28}^{56}\mathrm{Ni} \rightarrow {}_{27}^{56}\mathrm{Co} \rightarrow {}_{26}^{56}\mathrm{Fe}$.
In our SN model the mass of synthesized ${}^{56}\mathrm{Ni}$ is about 40\% of the mass of the exploded WD; i.e. $0.6\;M_{\odot}$. The energy released per ${}^{56}\mathrm{Ni}$ nuclear decay is a few MeV (see values in Table~\ref{tab:rad_decay} in Appendix~\ref{sec:decay}), we compute the total energy deposition in our SN model to be $1.1\times10^{50}$~erg, that is less than 8\% of the SN kinetic energy. This may produce significant heating of the ejecta (see the discussion in \cite{Woosley2007}), but cannot alter the overall dynamics of the SNR. And the nickel core is located deep inside the ejecta, it is barely reached by the RS at $t=500$~yr, so we do not expect it can strongly impact the morphology of the shocked region.\footnote{The situation may be different for a core-collapse SNR. Compare against the simulations by \cite{Wongwathanarat2015}, that show huge pillars of nickel growing toward the CD.} To check this, we included heating from radioactive decay in our SNR simulation, under the assumption of local deposition of all the energy released (in photons and in the kinetic energy of positrons), which constitutes an extreme case. See Appendix~\ref{sec:decay} for more details. We observed no significant differences on the projected maps and power spectra, so that our above results were not affected. A~more precise modeling would require hydrodynamic simulations coupled with radiative transfer, which is beyond the scope of this work.

\paragraph{Effect of the magnetic field.} 
Although the magnetic field is not expected to be strong enough to alter the development of young SNRs, and the key driver of the RTI is the deceleration of the ejecta, the RTI growth rate may be affected by the magnetization of the ejecta at the CD. From linear perturbation theory, a magnetic field tangential to the interface slows the growth of the instability due to the tension of the field. The instability is actually suppressed for wavelengths smaller than some critical value that is proportional to the energy density of the field \citep{Jun1995}. 
If the RTI from the SNR phase is suppressed, presumably it would be easier to see the contribution from the SN phase.
The development of the RTI into the non-linear stage in a strong magnetic field has been studied by \cite{Stone2007} in 3D, and including particle acceleration by \cite{Schure2009} in 2D. We~defer the use of MHD simulations to future work. 

\paragraph{Effect of particle acceleration.} 
Particle acceleration is expected to happen at the blast wave of SNRs via the mechanism of diffusive shock acceleration (DSA), and if particle acceleration is as efficient as it is believed to be, so that SNRs are significant producers of Galactic cosmic rays, then the fraction of the SN energy channeled into energetic particles may reach 10\% or more, which will impact the dynamics of the SNR (e.g. \cite{Ferrand2010} and references therein). 
The presence of energetic particles increases the compressibility of the fluid (lower effective adiabatic index), so the general effect is to make the shocked region more compact, meaning both narrower and denser. This effect is measurable and probably observed in Tycho's SNR \citep{Warren2005}, although it is not a huge modification. 

We believe that particle acceleration alone cannot strongly affect our conclusions regarding the morphology of the CD. 
The simulated X-ray maps produced by \cite{Ferrand2012} do not show obvious differences in the size of the structures present inside the SNR (the effect of particle acceleration is more visible in a differential comparison of the emission at different energy bands, that probe different levels of back-reaction). 
In \cite{Warren2013} the limited effect of lowering the adiabatic index on the power spectrum of density fluctuations can be seen in their Figure~6, at their final simulation time $t=2\;t_\mathrm{ch}=895$~yr that is almost twice our final time. They compare the standard case $\gamma=5/3$ with the extreme cases of $\gamma=4/3$ and $\gamma = 6/5$, which model the effect of the acceleration and escape of the relativistic particles (these simulations were made under the simplifying assumptions of reducing the adiabatic index everywhere in space and at all times). 

Regarding the shocks, a reduced distance between the CD and the FS will make it easier for the RT fingers to reach the FS, so that the morphology of the FS may contain some echo of the CD structure. Acceleration at the shock will not impact the shock surface by itself, unless the injection efficiency varies significantly on small scales all along the shock surface (observations of Tycho's SNR do not show evidence for this). The FS is more likely to be affected by other, external factors. As for the RS, efficient acceleration is not expected to occur in general, see the discussion in \cite{Ellison2005} (and there is no evidence for it in Tycho's SNR). 

\subsection{Implications for observations}
\label{sec:disc-obs}

Finally we compare our results with observations, especially of the young SNR Tycho (G120.1+1.4 = SN~1572, $447$~yr), which is the closest to our simulations.\footnote{
This SNR seems to be the best candidate to apply our findings, given its age and globally simple appearance. Other young type~Ia Galactic SNRs include G1.9+0.3, the youngest one known ($\simeq 150$~yr, \cite{Borkowski2013}); Kepler (G4.5+6.8 = SN~1604, $415$~yr), with a more complicated morphology presumably from the progenitor's mass loss history \citep{Reynolds2007}; and SN 1006 (G327.6+14.6, $1013$~yr), with a characteristic bilateral morphology from the orientation of the magnetic field \citep{Winkler2014}). Other young SN Ia SNRs in the LMC are SNR 0519-69.0 ($\simeq 450$~yr, \cite{Kosenko2010}) and SNR 0509-67.5 ($< 1000$~yr, \cite{Warren2004}).}
The most relevant energy band for this study is the X-ray band, which reveals the shocked matter, and for which we have instruments with sufficient angular resolution to study the detailed morphology. For Tycho's SNR, we refer the reader to the Chandra observations presented in \cite{Warren2005} and in \cite{Williams2017}.

Simulations of this SNR by \cite{Ferrand2012} and \cite{Warren2013}, that were made from spherically symmetric initial conditions, have reproduced the fleecy aspect of the inner ejecta, but have both failed to produce structures on scales as large as seen on the X-ray maps. \cite{Warren2013} noted that their power spectrum is significantly different from observations by \cite{Warren2005}, who found that the peak for the angular fluctuations of the CD (as seen in projection) occurs for $\ell = 6$. Our 1Di simulations confirm this: it appears that simulations made with different codes and both assuming spherical ejecta are in agreement with each other, and are in disagreement with X-ray observations of Tycho's SNR. And by lifting the assumption of spherically symmetry, our new work provides an explanation for this difference: the power observed at low~$\ell$ does not come from the RT growth but from the SN explosion. 
More precisely, we have observed in Section~3 that the contribution to the morphology of the CD from the SN phase is visible at low angular wavenumber~$\ell<10$, and decreasing over time, while the contribution from the SNR phase is visible at larger~$\ell$, increasing over time as well as shifting to lower~$\ell$. Interestingly, around 500~yr (about the age of Tycho's SNR) it so happens that the two contributions nicely match on the spectrum plot (last panel of Figure~\ref{fig:healpix_CD}): the power spectrum looks like a single distribution, just extending over a wider range of $\ell$ than would be expected from RTI alone. This could easily be confused with an enhanced RTI growth. And this actually looks like what is being observed in Tycho's SNR. 

\begin{figure}[t!]
\centering
\includegraphics[width=1\textwidth]{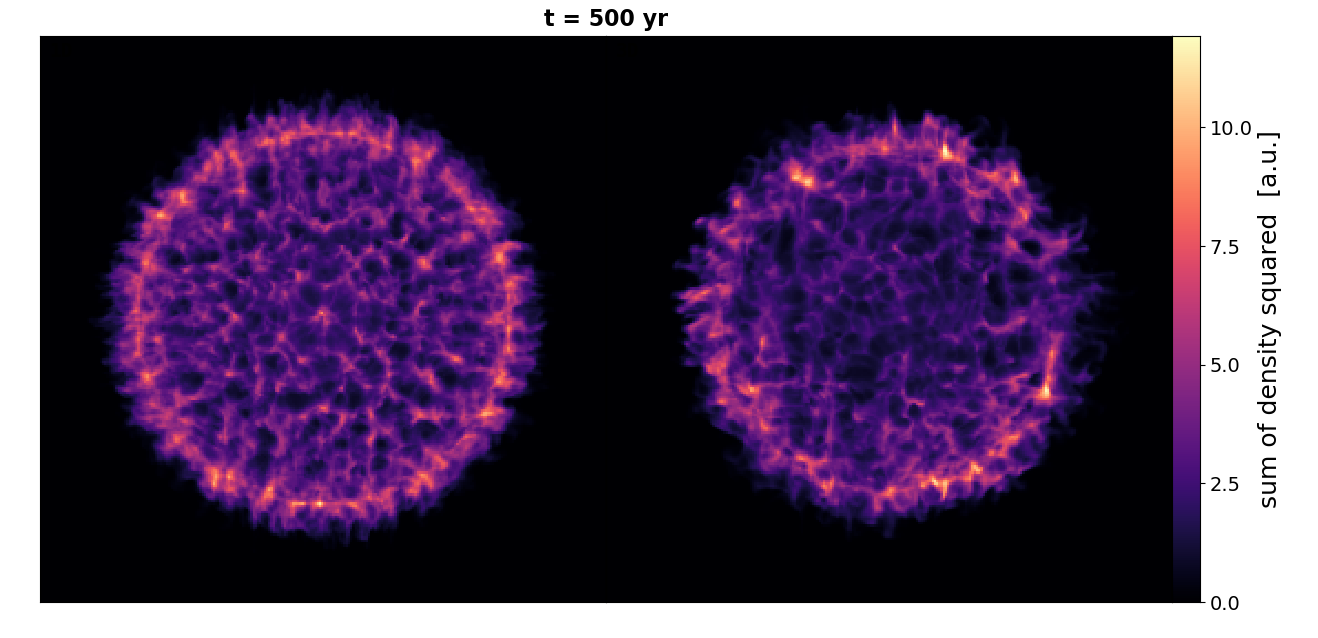}
\caption{\label{fig:em-2D}
Maps of the sum along an axis of the simulation cube of the mass density squared for the shocked ejecta, which is a proxy for the thermal emission as can be observed in X-rays. 
The left side shows the case of spherically symmetric ejecta (effectively 1D initial conditions), while the right side shows the case of asymmetric ejecta (fully 3D initial conditions). The maps are shown at an age of $t=500$~yr, close to the age at which Tycho's SNR is observed.}
\end{figure}

In Figure~\ref{fig:em-2D} we show an approximation of the thermal emission from the shock-heated ejecta, which scales as the square of the density and is observed summed up in projection along the line of sight. The more realistic SN model (used on the right) does produce a more realistic-looking mock SNR map, with a limb that shows more pronounced gaps and peaks like on Tycho's X-ray image. The emission from the shocked plasma actually depends on the electronic density, the electronic temperature, and the ionization states of all the species present, which for young SNRs depend on the history of the shocked material since the crossing of the shock (see \cite{Ferrand2012} and references therein). In a forthcoming paper we will calculate X-ray maps and present a dedicated analysis of Tycho's SNR data. 

 For the case of Tycho's SNR, \cite{Williams2017} compared 3D simulations made from smooth initial ejecta profiles and from clumpy initial ejecta profiles (with a characteristic scale set by hand). Measuring the deceleration parameter in ejecta knots, they were not able to tell apart the two models using their X-ray observations, although visual analysis of the 2D projected images shows clear differences in the morphology of the SNR depending on the initial conditions (compare their Figure~10 to our Figure~\ref{fig:em-2D}). \cite{Sato2019} use the \emph{genus statistics}, a topological method, to quantify these differences. They find that the observed morphology of Tycho better matches the case of clumpy initial ejecta. This comforts our view that the SN explosion has a visible imprint on this SNR.

In this work we deliberately assumed a uniform ISM, so as to reveal the contribution from the SN. The power in the fluctuations of the RS and FS is continuously decaying over time, and after hundreds of years is very small, probably too difficult to detect. From observations of young SNRs it is relatively easy to delineate the FS, which is traced by the synchrotron emission of accelerated electrons (see again \cite{Warren2005} as well as \cite{Cassam-Chenai2007} for the X-ray observations of Tycho's SNR, and \cite{Ferrand2014} for the corresponding simulations). But by that age and on those scales the FS will most probably be impacted by inhomogeneities in the ambient medium, which are not included in our simulations.\footnote{For an example applied to Tycho's SNR, see the recent work by \cite{Fang2018}, who considered the impact of a stellar outflow from the progenitor system to explain irregularities along the outer shape of the remnant.} 
The RS, that runs inside the ejecta, will not be as immediately affected as the FS by the ISM. However it is in general more difficult to locate from observations, and as a result comparisons will be subject to some degree of interpretation (e.g., for the case of Tycho, relying on a particular emission line, the Fe K$\alpha$ line, as suggested by \cite{Warren2005}).
So we expect it will be more difficult to draw firm conclusions from studies of the SNR shocks. Density fluctuations in the shocked ejecta (as delineated by the CD) seem to contain more exploitable information.
We note that, as time goes by, the RS will reveal in X-rays the ejecta located deeper and deeper inside the ejecta. At the current age of Tycho, the core of the ejecta is invisible in X-rays, and unfortunately the cold ejecta have not been detected in a type Ia SNR at other wavelengths, be it in radio (molecular lines), infrared (dust) or gamma (radioactive lines). Although such detections are challenging, they would open a new window to the SN and SNR physics, that we could take advantage of using our simulations.

\section{Conclusion}
\label{sec:conclusion}

In this paper we have studied the imprint of a SN explosion on the morphology of the resulting SNR, by the means of 3D hydrodynamics simulations. We have concentrated on the thermonuclear case (for type Ia SNe), and used the recent N100 model from \cite{Seitenzahl2013}, a delayed detonation (DDT) of a Chandrasekhar mass white dwarf, that we put in a low-density uniform medium thought to be a good first approximation for Tycho's SNR. 
Interestingly, using a realistic 3D SN model leads to larger scale and more irregular structures, which were not seen in SNR simulations made from (semi-)analytical SN models, and which better match X-ray observations of Tycho's SNR.
To quantify the effect, we have extracted the time-dependent surfaces of the forward shock, contact discontinuity, and reverse shock, and analyzed their angular scale structure (see Figures~\ref{fig:healpix_CD}, \ref{fig:healpix_RS}, \ref{fig:healpix_FS}). By comparing, as initial conditions, the actual three-dimensional distribution of the ejecta obtained from the SN simulation with an angularly smoothed version of it, we have shown that morphological signatures of the explosion can be seen clearly in the first hundred years, and can still be detected after a few hundred years. For the forward and reverse shocks the angular power from the SN only decreases over time, and for the forward shock may be further erased by inhomogeneities in the ambient medium. For the CD the SN power decreases over time while the SNR power increases, due to the Rayleigh-Taylor instability. Thanks to the expansion in spherical harmonics we have devised a way to separate the contributions of the SN and SNR phases. For the CD they are dominant respectively above and below scales of about $20\deg$; in terms of total power the SNR phase takes over at around 200 yr (see Figure~\ref{fig:power_time}). For our Tycho-like SNR the two contributions to the angular spectrum, from the SN phase and from the SNR phase, are found to have similar power at the current observed age, which mimics a faster RTI growth (see power spectrum in Figure~\ref{fig:healpix_CD}~(c)). We suggest this is what is actually observed in Tycho's SNR.

The obvious extension of this work is to re-do our simulation and analysis with different SN models. We have access to a grid of DDT models, as well as a grid of pure deflagration models. We can also test sub-Chandrasekhar mass Type Ia supernova models, as well as other channels for Type~Ia SNe (double degenerate scenario). By investigating all these different models, we will be able to assess the range of behaviors that may be observed for the case of type Ia SN(R)s, in terms of the surviving asymmetry of the ejecta. We note that the N100 model used here is a 3D explosion model having a quite high degree of spherical symmetry. Other 3D DDT models, and especially merger models, can exhibit much greater asymmetry. We also note that in the single-degenerate case there is a surviving companion, which would also interact with the ejecta \citep[e.g.][]{Liu2012,Pakmor2008}, and could perhaps lead to additional structure in the remnant. 
With this program, we can determine if and how the SN signatures that we have established in this paper can be used to discriminate between different thermonuclear SN theories, which is an important step to be able to use SNRs as probes of the explosion physics.

\acknowledgments
This work was supported in part by RIKEN Interdisciplinary Theoretical and Mathematical Sciences Program (iTHEMS). This work was supported in part by a RIKEN pioneering project ``Extreme precisions to Explore fundamental physics with Exotic particles (E3-Project)''.
The work of FKR was supported by the Klaus Tschira Foundation and by the German Research Foundation (DFG) via the Collaborative Research Center SFB 881 ``The Milky Way System''. IRS was supported by Australian Research Council Grant FT160100028.
We thank the anonymous referee for comments that helped clarify the text.

\facilities{Simulations were performed on the iTHEMS clusters at RIKEN.}

\software{HEALPix \citep{Gorski2005}, SciPy \citep{Jones2001}, Matplotlib \citep{Hunter2007}}

\newpage
\appendix

\section{Co-expansion transformation}
\label{sec:comoving}

We detail here the transformation that makes the computational grid co-expand with the SNR, without any need to add grid cells or refinement layers. The technique is commonly used for cosmological simulations, it was applied to SNRs by \cite{Fraschetti2010}, following \cite{Poludnenko2007}.\footnote{The supernova explosion itself was computed on an expanding grid, using a different ``moving grid'' technique \citep{Ropke2005}.} Everywhere in this section \emph{tilde} quantities refer to transformed quantities, i.e. after the change of variable.

\subsection{The general transformation}

We need to define the transformation for the independent space, time, and mass dimensions. 
Distance~$r$, time~$t$, and mass density~$\rho$ are transformed as
\begin{eqnarray}
\mathbf{\tilde{r}} & = & a^{-1}\:\mathbf{r} \label{eq:r_tilde}\\
\mathrm{d}\tilde{t} & = & a^{-1-\beta}\:\mathrm{d}t \label{eq:dt_tilde}\\
\tilde{\rho} & = & a^{\alpha}\:\rho \label{eq:rho_tilde}
\end{eqnarray}
where $\alpha$ and $\beta$ are free parameters, set below. 
This in turns implies the velocity transformation
\begin{equation}
\mathbf{\tilde{u}} = \frac{\mathrm{d}\mathbf{\tilde{r}}}{\mathrm{d}\tilde{t}} = a^{\beta}\left(\mathbf{u}-H\mathbf{r}\right) \label{eq:u_tilde}
\end{equation}
with expansion factor (``Hubble factor'')
\begin{equation}
H=\frac{1}{a}\frac{\mathrm{d}a}{\mathrm{d}t}=\frac{\mathrm{d}\ln a}{\mathrm{d}t}
\label{eq:H}
\end{equation}
and also
\begin{eqnarray}
\tilde{P} & = & a^{\alpha+2\beta}\:P \label{eq:P_tilde}\\
\tilde{E} & = & a^{\alpha+2\beta}\:E \label{eq:E_tilde}\\
\tilde{\epsilon} & = & a^{2\beta}\:\epsilon \label{eq:eps_tilde}
\end{eqnarray}
for pressure~$P$, total energy~$E$, energy per unit mass~$\epsilon$.

The Euler equations (here written in conservative form) become
\begin{equation}
\frac{\partial\tilde{\rho}}{\partial\tilde{t}}+\mathbf{\tilde{\nabla}.}\left(\tilde{\rho}\tilde{\mathbf{u}}\right)  =  (\alpha-\nu)\tilde{H}\;\tilde{\rho} \label{eq:Euler-mass}
\end{equation}
\begin{equation}
\frac{\partial\tilde{\rho}\tilde{\mathbf{u}}}{\partial\tilde{t}}+\mathbf{\tilde{\nabla}.}\left(\tilde{\rho}\mathbf{\tilde{u}\tilde{u}}\right)  =  (\alpha-\nu+\beta-1)\tilde{H}\:\tilde{\rho}\mathbf{\tilde{u}}-\mathbf{\tilde{F}} \label{eq:Euler-momentum}
\end{equation}
\begin{equation}
\frac{\partial\tilde{E}}{\partial\tilde{t}}+\mathbf{\tilde{\nabla}.}\left(\left(\tilde{E}+\tilde{P}\right)\tilde{\mathbf{u}}\right)  =  \left(\alpha-\nu+2(\beta-1)\right)\tilde{H}\:\tilde{E}-\mathbf{\tilde{F}.\tilde{u}}+\left(\frac{2}{\gamma-1}-\nu\right)\tilde{H}\;\tilde{P} \label{eq:Euler-energy}
\end{equation}
where $\nu$ is the number of spatial dimensions (for our case $\nu=3$), $\tilde{H}$ is the comoving expression of the expansion factor:
\begin{equation}
\tilde{H}(t) = a^2 H(t)\;,
\label{eq:H_tilde}
\end{equation}
and we have introduced the (comoving) inertial force 
\begin{equation}
\mathbf{\tilde{F}}\left(\tilde{\mathbf{r}},\tilde{t}\right)=\tilde{f}\left(\tilde{t}\right)\:\tilde{\rho}\left(\tilde{\mathbf{r}},\tilde{t}\right)\:\mathbf{\tilde{r}} 
\label{eq:F_tilde}
\end{equation}
where
\begin{equation}
\tilde{f}=-\left(\frac{\mathrm{d}\tilde{H}}{\mathrm{d}\tilde{t}}-\beta\tilde{H}^{2}\right)=-\left(a^{2\beta+1}\frac{\mathrm{d}^{2}a}{\mathrm{d}t^{2}}\right) 
\label{eq:f_tilde}
\end{equation}
In order to keep the Euler equations as close as possible from their original conservative form, we choose $\alpha=\nu=3$ and $\beta=1$. Then, for a fluid with adiabatic index $\gamma=5/3$ (no internal degrees of freedom), the only extra terms are the ones involving the force~$\mathbf{\tilde{F}}$. This stems from the fact that we are now working in a non-inertial frame. 

\subsection{Application to an expanding SNR}

The expansion law of a SNR is commonly written as a power-law
\begin{equation}
a(t)=a_{\star}\left(\frac{t}{t_{\star}}\right)^{\lambda}
\label{eq:a(t)=t^lambda}
\end{equation}
where $a_{\star}$ is the size at some reference time $t_{\star}$ (for us the physical start time~$t_0$ of the SNR simulation), for some constant value of index~$\lambda$.
Self-similar phases obey such a relation: in the initial free expansion phase $\lambda=1$, the Chevalier solution for the ejecta-dominated phase with power-law profiles of the ejecta (index~$n$) and of the ISM (index~$s$) gives $\lambda=(n-3)/(n-s)$, in the later Sedov phase conservation of energy implies $\lambda=0.4$.
In that case we have simple expressions for the derivatives
\begin{eqnarray}
\frac{\mathrm{d}a}{\mathrm{d}t} & = & \lambda\frac{a(t)}{t} \label{eq:da/dt}\\
\frac{\mathrm{d}^{2}a}{\mathrm{d}t^{2}} & = & \lambda(\lambda-1)\frac{a(t)}{t^{2}} \label{eq:d2a/dt2}
\end{eqnarray}
and we can integrate the relation between physical and comoving time:
\begin{equation}
\frac{\tilde{t}-\tilde{t}_{0}}{\tilde{t}_{\star}}=
\begin{cases}
\frac{1}{1-2 \lambda}\left(\left(\frac{t}{t_{\star}}\right)^{1-2 \lambda}-\left(\frac{t_{0}}{t_{\star}}\right)^{1-2 \lambda}\right) & \lambda\neq\frac{1}{2} \\
\ln\left(\frac{t}{t_{\star}}\right)-\ln\left(\frac{t_{0}}{t_{\star}}\right)=\ln\left(\frac{t}{t_{0}}\right) & \lambda=\frac{1}{2} 
\end{cases}
\label{eq:t_tilde(t)}
\end{equation}
where $\tilde{t}_{0}$ is the arbitrary starting value of the comoving time and $\tilde{t}_{\star}=t_{\star}/a^2_{\star}$.

When the scale factor has more complicated time dependence,\footnote{For an analytic description of the continuous evolution in time of a (non-radiative) SNR, see \cite{Tang2017}.} we can always define the instantaneous index
\begin{equation}
\lambda(t) = \frac{da/a}{dt/t} = \frac{d\ln(a)}{d\ln(t)}
\label{lambda(t)}
\end{equation}
and we still have
\begin{equation}
H(t) = \frac{1}{a}\frac{da}{dt} = \frac{\lambda(t)}{t}
\label{H(t)}
\end{equation}
and in comoving coordinates relations~(\ref{eq:H_tilde}) and~(\ref{eq:f_tilde}).
We now write
\begin{equation}
\frac{1}{a} \frac{d^2a}{dt^2} = \frac{\lambda(t)(\lambda(t)-1) + \mu(t)}{t^2}
\label{eq:1/a d2a/dt2}
\end{equation}
with 
\begin{equation}
\mu(t) = \frac{d\lambda(t)}{d\ln(t)}\;,
\label{eq:mu(t)}
\end{equation}
which generalizes Eq.~(\ref{eq:d2a/dt2}). 

In practice the simulation is done in two passes. In~the first pass we use a fixed $\lambda\lesssim 1$ and an oversized box, and record the average position $r_\mathrm{CD}$ of the contact discontinuity over time. We define $a(t)=r_\mathrm{CD}(t)/r_\mathrm{CD}(t_0)$, and integrate the physical time $t$ to compute the comoving time~$\tilde{t}$ according to Eq.~(\ref{eq:dt_tilde}) (from $\tilde{t}=0$ when $t=t_0$). We also compute the derivatives of $a(t)$ w.r.t. $\ln(t)$: $\lambda(t)$ and $\mu(t)$, needed to compute $\tilde{H}$ and~$\mathbf{\tilde{F}}$, from a spline interpolator of $a(t)$ (we used \texttt{scipy.interpolate.spalde}). In~the second pass we use the tabulated scaling law to define the transformed quantities at each time-step.

Finally we note that a side-effect of having a quasi-stationary SNR in the computational grid is the development of a numerical instability at the FS front (carbuncle instability, e.g. \cite{Quirk1994}). Compared to our previous studies, this is far less of a concern when using a 3D SN model, since the FS is never perfectly spherical. Besides, the high-spatial frequency noise generated (at the cell scale) is irrelevant to the effects discussed here. 


\section{Analysis of the wave structure}
\label{sec:waves-analysis}

In this section we detail the practical aspects, that are specific to our simulation setup, of the extraction and analysis of surfaces inside the SNR.

\subsection{Extracting and recording wave fronts}

During runtime, at every hydro time step, we look for the position of the waves inside the simulation box. These interfaces are located by finding where some quantity crosses a threshold, along each of the three spatial dimensions. 
For the CD, this is the ejecta fraction $f$, we set the threshold to $f=0.5$. Varying this threshold between 0.1 and 0.9 causes only minor changes on the maps and spectra. 
For shocks, the most obvious marker is the pressure jump, the threshold on $\log(P)$ is set to be the arithmetic mean of the lowest (ambient) and highest (shocked) pressure in the box. Numerically the width of the shocks is only of a few cells, so the exact value of the threshold is not critically important, as long as it is applied consistently in time and across simulation runs. The distinction between the FS and the RS can readily be made knowing the ejecta fraction.
Working on a Cartesian grid, we thereby collect an (unordered) set of faces that are boundaries between grid cells. We also record their orientation, obtained from the gradient of the quantity~$f$ or~$P$, for visualization purpose. 
This scheme ensures that we extract a contiguous surface for each wave. We note that the surface of the CD may present folds, meaning that its radius is a multi-valued function along a single ray from the centre. 

The location of each wave (radial distance from the SN center to the face center) is then projected on the surface of a sphere, which is tesselated using the HEALPix scheme \citep{Gorski2005}. We adopt the RING ordering. The HEALPix tessellation is controlled by a resolution parameter $N_\mathrm{side} = 2^k$ where $k$ is an integer, the total number of pixels being $N_\mathrm{pixel} = 12 N_\mathrm{side}^2 = 12\times2^{2k}$. The resolution of the Cartesian grid on the other hand is controlled by the maximum level number $k_\mathrm{max}$ of the RAMSES octree, so that the number of cells on the finest level is $2^{k_\mathrm{max}}$ along each of the 3~dimensions. We set $k=k_\mathrm{max}-1$ so that the angular resolution of the HEALPix map be appropriate for the Cartesian grid (of course the individual Cartesian faces will project at a variety of angles). We tried varying the parameter~$k$ for a given $k_\mathrm{max}$, and checked that differences between maps of varying HEALPix resolution are modest, and would not affect our diagnostics. Our baseline is $k_\mathrm{max}=8$ so that we have $256^3=16,777,216$ cells (and $6\times 256^2=393,216$ outer faces), and $k=7$ so that for the HEALPix maps $N_\mathrm{side} = 128$ and $N_\mathrm{pixel} = 196,608$. 
 
To project the faces that make up the wave surfaces, we break them down recursively into four smaller faces, and assign their radius to the HEALPix pixels in the direction of all the leaf sub-faces. Four recursion steps are sufficient to obtain a converged projected map. This high-resolution projection is done only when outputting the wave data to disk (which is done independently of the grid outputs). 
Furthermore, for the CD, which might have a complicated morphology in 3D, we tried two projection schemes: collecting all the faces, thereby averaging over folds, or collecting only the outermost faces in a given direction. The two schemes result in small differences that are visible on the maps, but barely affect the spectra defined below. We present here maps made using the first scheme.

\subsection{Quantifying asymmetries with spherical harmonics expansion.}

At this stage, we have a representation of the morphology of each of the wave fronts (RS, CD, FS) as 
a function on the sphere, $r(\theta,\phi)$, the radius at which this wave front is found along direction $(\theta,\phi)$. We normalize this function as $R=(r-\langle r\rangle)/\langle r\rangle$ where $\langle r\rangle$ is the average value over all angles, so that we are working with relative fluctuations in radius. To quantify these fluctuations, we expand $R(\theta,\phi)$ in spherical harmonics:
\begin{equation}
R\left(\theta,\phi\right) = \sum_{\ell=0}^{\infty}\sum_{m=-\ell}^{+\ell}a_{\ell}^{m}Y_{\ell}^{m}\left(\theta,\phi\right)
\end{equation}
where $Y_{\ell}^{m}$ is the Laplace spherical harmonic of degree~$\ell\geq0$ and order~$m\in\left[-\ell,+\ell\right]$, and the expansion coefficients $a_{\ell}^{m}$ are found by projection of~$R$ on the orthogonal basis of the $Y_{\ell}^{m}$ functions. This is done in post-processing on the HEALPix map, using the Python package \texttt{healpy}.\footnote{\url{https://healpy.readthedocs.io}}
The degree~$\ell$ determines the spatial frequency of the function over the sphere, which can be estimated as $\pi/\ell$.

The angular power spectrum is defined as
\begin{equation}
C_{\ell}=\overline{\left|a_{\ell}^{m}\right|^{2}}=\frac{1}{2\ell+1}\sum_{m=-\ell}^{+\ell}\left|a_{\ell}^{m}\right|^{2}\,,
\end{equation}
and Parseval's theorem reads
\begin{equation}
\int_{\theta=0}^{\pi}\int_{\phi=0}^{2\pi}|R\left(\theta,\phi\right)|^2\,\mathrm{d}\Omega
=\sum_{l=0}^{\infty}\sum_{m=-\ell}^{+\ell}\left|a_{\ell}^{m}\right|^{2}=\sum_{l=0}^{\infty}\left(2\ell+1\right)C_{\ell}
\end{equation}
so that the contribution of scale~$\ell$ to the total variance of~$R$ over the sphere is $\left(2\ell+1\right)C_{\ell}/4\pi$. In all the figures we show the spectrum as a function of $\log \ell$ and therefore plot this latter quantity multiplied by~$\ell$, so that the grayed area be the variance of the function~$R$.\footnote{We note that the convention used in all CMB plots is neither of these, but $\ell(\ell+1)C_{\ell}/2\pi$, for historical reasons specific to cosmology (the Sachs-Wolfe effect).}


\section{Energy deposition from radioactive decay}
\label{sec:decay}

In this section we outline the basics of energy deposition from radioactive decay. The general formalism is taken from \cite{Jeffery1999}, see also \cite{Nadyozhin1994}.

The energy deposition per unit volume and time is noted $\rho\,\epsilon$ with a time-dependent energy production rate $\epsilon = C\,f(t)$, where $C$ is the energy generation rate coefficient and $f(t)$ is the fraction of species left at time $t$. 
For primary species: 
\begin{equation}
C = \frac{Q}{\mathrm{amu}\,A\,t_e}\;,
\end{equation}
\begin{equation}
f(t) = f(t_0) \exp\left(-\frac{t}{t_e}\right)\;,
\end{equation}
and for secondary species ($P$ designates the parent species): 
\begin{equation}
C = \frac{Q}{\mathrm{amu}\,A_P\,(t_e-t_{e,P})}\;,
\end{equation}
\begin{equation}
f(t) = f_P(t_0) \left(\exp\left(-\frac{t}{t_e}\right) - \exp\left(-\frac{t}{t_{e,P}}\right)\right),
\end{equation}
where $Q$ is the mean energy deposit per decay in photons or e$^+$/e$^-$ kinetic energy (the energy that can be absorbed by matter)
and $t_e$ is the $e$-folding time ($t_e=t_{1/2}/\ln(2)$ where $t_{1/2}$ is the half-life).

We are here interested in the (completely dominant) decay chain 
\begin{align}
&
{}_{28}^{56}\mathrm{Ni} \rightarrow {}_{27}^{56}\mathrm{Co} + \gamma + \nu_e\\
&
{}_{27}^{56}\mathrm{Co} \rightarrow \begin{cases}
{}_{26}^{56}\mathrm{Fe} + \gamma + \nu_e         & (81\%) \\
{}_{26}^{56}\mathrm{Fe} + \gamma + \nu_e + e^{+} & (19\%). \\
\end{cases}
\end{align}
So in the hydro code we add a source term for the internal energy $\rho \left(\epsilon_{\mathrm{Ni}}+\epsilon_{\mathrm{Co}}\right)$, where $\rho$ is the ejecta mass density and the initial fraction of ${}^{56}\mathrm{Ni}$ is known from the SN simulation. The relevant numerical values are compiled in Table~\ref{tab:rad_decay}. We indicate the total energy deposit as well as the part that goes into photons and kinetic energy of positrons, that may be converted into heat (excluding neutrinos, which nearly entirely escape). 

In our simple tests, we assume local energy deposition, without radiative transfer. Initially the optical depth is very large, caused by Compton scattering. The transition time between optically thick and optically thin periods is estimated by \cite{Jeffery1999} to be about 40~yr for their fiducial Type Ia SN (their Equation~(28)). For comparison, 97\% of the energy liberated from radioactive decay of ${}^{56}\mathrm{Ni}$ is deposited within the first year after the explosion. The total energy deposition for the N100 model is $1.1\times10^{50}$~erg, that is less than 8\% of the SN kinetic energy.

\begin{deluxetable}{lll}[h]
\tablehead{\colhead{} & \colhead{${}_{28}^{56}\mathrm{Ni}$}& \colhead{${}_{27}^{56}\mathrm{Co}$}}
\startdata
$t_{1/2}$ = half-life (days) & 6.077(12) & 77.27(3) \\
$t_e$ = $e$-folding time (days) & 8.767(12) & 111.48(4) \\
$Q$ total (MeV) & 2.135(11) & 4.566(2) \\
$Q$ photons+kinetic (MeV) & 1.729(17) & 3.74(4) \\
$C$ (erg.s$^{-1}$.g$^{-1}$) & $3.94(4)\times10^{10}$ & $7.27(7)\times10^9$
\enddata
\caption{\label{tab:rad_decay}
Data for the radioactive decay of ${}_{28}^{56}\mathrm{Ni}$ and ${}_{27}^{56}\mathrm{Co}$ \citep[from][]{Jeffery1999}.}
\end{deluxetable}


\newpage

\bibliographystyle{aasjournal}
\bibliography{references}

\end{document}